\documentclass[a4paper,11pt]{article}
\pdfoutput=1 

\usepackage{jheppub} 

\usepackage[T1]{fontenc} 

\usepackage{graphicx,epsfig,color}
\usepackage[english]{babel}
\usepackage{amssymb,amsfonts,amsmath}
\usepackage{verbatim}

\newcommand{\ord}[2]{{\ensuremath{{#1}^{(#2)}}}}
  \newcommand{\cN}{{\cal N}}

\def\op{\ord{p}{1}}
\def\opp{\ord{p}{2}}
\def\oppp{\ord{p}{3}}
\def\oe{\ord{\varepsilon}{1}}
\def\oee{\ord{\varepsilon}{2}}
\def\om{\ord{\mn}{1}}
\def\omm{\ord{\mn}{2}}
\def\oM{\ord{M}{1}}
\def\oMM{\ord{M}{2}}
\def\oA{\ord{A}{0}}

\def\oP{\ord{P}{0}}

\def\oH{\ord{H}{0}}
\def\oHH{\ord{H}{1}}
\def\oB{\ord{B}{0}}

\def\oQ{\ord{Q}{0}}

\def\oBi{\ord{B_i}{0}}
\def\oBBi{\ord{B_i}{1}}
\def\oBo{\ord{B_o}{0}}
\def\oBBo{\ord{B_o}{1}}
\def\oAi{\ord{A_i}{0}}
\def\oAAi{\ord{A_i}{1}}
\def\oAo{\ord{A_o}{0}}
\def\oAAo{\ord{A_o}{1}}

\def\oQi{\ord{Q_i}{0}}
\def\oQQi{\ord{Q_i}{1}}
\def\oQo{\ord{Q_o}{0}}
\def\oQQo{\ord{Q_o}{1}}
\def\oPi{\ord{P_i}{0}}
\def\oPPi{\ord{P_i}{1}}

\def\oHi{\ord{H_i}{0}}
\def\oHHi{\ord{H_i}{1}}
\def\oHo{\ord{H_o}{0}}
\def\oHHo{\ord{H_o}{1}}

\def\obo{\ord{b_o}{0}}
\def\obbo{\ord{b_o}{1}}

\def\oao{\ord{a_o}{0}}
\def\oaao{\ord{a_o}{1}}

\def\okl{\ord{\kel}{0}}
\def\okkl{\ord{\kel}{1}}

\def\on{\ord{\nu}{1}}
\def\onn{\ord{\nu}{2}}

\def\oR{\ord{R}{0}}
\def\oRR{\ord{R}{1}}

\def\kel{ {k_l^{\textup{el}}} }
\def\kmag{ {k_l^{\textup{mag}}} }

  \def\mn{{\mathfrak{m}}}
 
\newcommand{\be}{\begin{equation}} \newcommand{\ee}{\end{equation}}
\newcommand{\bea}{\begin{eqnarray}} \newcommand{\eea}{\end{eqnarray}}
\newcommand{\beann}{\begin{eqnarray*}}  \newcommand{\eeann}{\end{eqnarray*}}
\newcommand{\bfig}{\begin{figure}} \newcommand{\efig}{\end{figure}}
\newcommand{\ba}{\begin{array}} \newcommand{\ea}{\end{array}}
\newcommand{\bcen}{\begin{center}} \newcommand{\ecen}{\end{center}}
\newcommand{\btab}{\begin{tabular}} \newcommand{\etab}{\end{tabular}}

\newcommand{\bp}{\begin{Proposition}}   \newcommand{\ep}{\end{Proposition}}
\newcommand{\bt}{\begin{Theorem}}   \newcommand{\et}{\end{Theorem}}
\newcommand{\bl}{\begin{Lemma}}     \newcommand{\el}{\end{Lemma}}
\newcommand{\bc}{\begin{Corollary}} \newcommand{\ec}{\end{Corollary}}
\newcommand{\tred}[1]{#1}

\title{Holographic compact stars meet gravitational wave constraints}

\author[a]{Eemeli Annala}
\author[b]{Christian Ecker}
\author[c]{Carlos Hoyos}
\author[a]{Niko Jokela}
\author[c,d]{David Rodr\'{\i}guez Fern\'andez}
\author[a]{Aleksi Vuorinen}

\affiliation[a]{Department of Physics and Helsinki Institute of Physics\\
P.O.~Box 64, FI-00014 University of Helsinki, Finland}
\affiliation[b]{Institut f\"ur Theoretische Physik, Technische Universit\"at Wien \\
Wiedner Hauptstr.~8-10, A-1040 Vienna, Austria}
\affiliation[c]{Department of Physics, Universidad de Oviedo\\
Avda.~Calvo Sotelo 18, ES-33007 Oviedo, Spain}
\affiliation[d]{Institute for Theoretical Physics and Astrophysics, University of W\"urzburg\\
97074 W\"urzburg, Germany}

\emailAdd{eemeli.annala@helsinki.fi}
\emailAdd{christian.ecker@tuwien.ac.at}
\emailAdd{hoyoscarlos@uniovi.es}
\emailAdd{niko.jokela@helsinki.fi}
\emailAdd{david.rodriguez@physik.uni-wuerzburg.de}
\emailAdd{aleksi.vuorinen@helsinki.fi}

\abstract{\small We investigate a simple holographic model for cold and dense deconfined QCD matter consisting of three quark flavors. Varying the single free parameter of the model and utilizing a Chiral Effective Theory equation of state (EoS) for nuclear matter, we find four different compact star solutions: traditional neutron stars, strange quark stars, as well as two non-standard solutions we refer to as hybrid stars of the second and third kind (HS2 and HS3). The HS2s are composed of a nuclear matter core and a crust made of stable strange quark matter, while the HS3s have both a quark mantle and a nuclear crust on top of a nuclear matter core. For all types of stars constructed, we determine not only their mass-radius relations, but also tidal deformabilities, Love numbers, as well as moments of inertia and the mass distribution. We find that there exists a range of parameter values in our model, for which the novel hybrid stars have properties in very good agreement with all existing bounds on the stationary properties of compact stars. In particular, the tidal deformabilities of these solutions are smaller than those of ordinary neutron stars of the same mass, implying that they provide an excellent fit to the recent gravitational wave data GW170817 of LIGO and Virgo. The assumptions underlying the viability of the different star types, in particular those corresponding to absolutely stable quark matter, are finally discussed at some length.}

\preprint{{\small FPAUO-17/15, HIP-2017-28/TH}}


\begin{document}
\maketitle


\section{Introduction}

\begin{figure}[ht!]
\begin{center}
\includegraphics[width=0.45\textwidth]{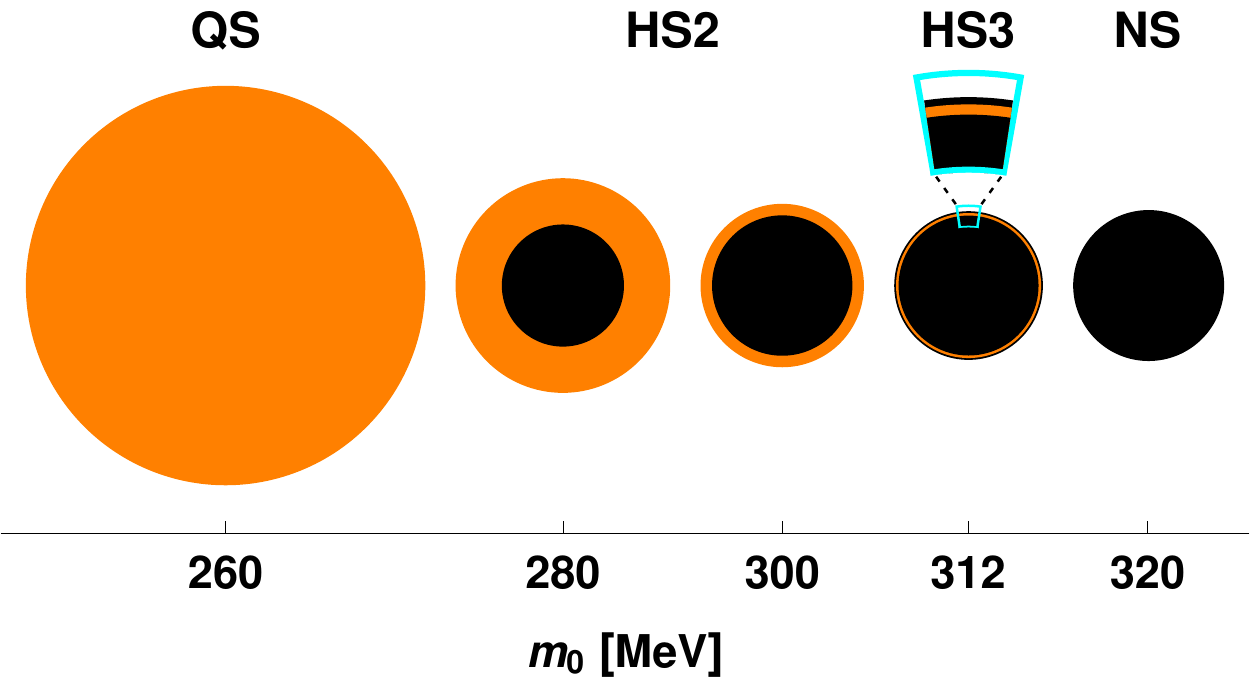}
\caption{An illustration of the structure of two-solar mass stars obtained for different values of $m_0$. The chosen cases correspond to a QS, two different hybrid stars of type HS2, one HS3, and one ordinary NS. The orange (black) color represents quark (nuclear) matter, while the radii of the different circles are proportional to the actual sizes of the corresponding regions inside the stars. In Fig.~\ref{fig:MRmulti2}, these five stars are denoted by small crosses on the corresponding $M$--$R$ curves.  
}\label{fig:MRmulti}
\end{center}
\end{figure}

The nature and properties of compact stars is a topic of active research both on the observational and theoretical sides \cite{Lattimer:2004pg}. The standard picture is that all stars with densities comparable to the nuclear matter saturation density $n_s$ are neutron stars (NS), composed of hadronic matter of increasing density, or hybrid stars (HS) that in addition contain deconfined quark matter in their inner cores (this class also includes so-called twin stars). This scenario is based on the assumption of nuclear matter being absolutely stable in vacuum, i.e., that it has a lower energy per baryon ratio at zero pressure than quark matter. Albeit a highly plausible assumption --- after all, we know from observations that at least most of the compact stars detected so far appear to have masses and radii in the range predicted for NSs --- the case for stable three-flavor quark matter and so-called strange quark stars (QS) has not been settled yet \cite{Itoh:1970uw,Bodmer:1971we,Terazawa:1979hq,Farhi:1984qu,Witten:1984rs}. In particular, a scenario with two separate families of compact stars with different mass-radius ($M$--$R$) branches, one corresponding to NSs or HSs and the other to QSs, remains viable \cite{Fraga:2001id,Weber:2004kj,Postnikov:2010yn,Drago:2013fsa}. Inherent in these rather exotic proposals is the nontrivial assumption that finite-size effects resolve problems related to, e.g., unobserved quark matter halos being formed around atomic nuclei.

On the theory side, the difficulty in excluding the existence of absolutely stable quark matter is related to the fact that no robust first principles tools exist for studying this phase of Quantum Chromodynamics (QCD) at moderate energy densities  \cite{Brambilla:2014jmp}, including quark matter in its strongly coupled regime just above the deconfinement transition density. With the Sign Problem impeding lattice studies utilizing Monte-Carlo simulations \cite{deForcrand:2010ys}, and weak coupling methods being restricted to the ultrahigh-density regime \cite{Freedman:1976ub,Vuorinen:2003fs,Kurkela:2009gj,Kurkela:2016was}, the options that remain include investigating simplified models of QCD (see, e.g., \cite{Buballa:2003qv}) or deforming the theory to allow for a nonperturbative solution even at strong coupling. A prime example of the latter approach is naturally the gauge/gravity duality \cite{Maldacena:1997re,Gubser:1998bc,Witten:1998qj}, which in particular allows the description of a class of strongly coupled theories with flavor degrees of freedom \cite{Karch:2002sh}. 

Within the past two decades, the gauge/gravity duality has been frequently applied to the description of the quark-gluon plasma (QGP) produced in heavy ion collisions (see, e.g., \cite{CasalderreySolana:2011us,Brambilla:2014jmp} for reviews). The models that are typically considered in this context are dual to theories that differ from QCD in a number of ways (e.g.~by exhibiting supersymmetry and conformal invariance), and are furthermore studied in their large-$N$ and infinitely strongly coupled limits. Despite these unphysical features, at nonzero temperatures the holographic systems exhibit universal properties that qualitatively match with those measured in heavy ion collisions, including in particular a fast thermalization rate as well as a hydrodynamic expansion with an almost perfect fluid behavior. In fact, experimental estimates of the shear viscosity of the QGP fall remarkably close to the value predicted by holographic models \cite{Kovtun:2004de}, which has prompted further studies of the properties of the QGP by means of the duality.

It is worth highlighting that the holographic duality is able to capture the qualitative and in some cases even quantitative properties of the QGP through the study of theories that have vacuum properties very different from QCD. An obvious question then arises concerning whether this behavior is specific to high temperatures, or if a similar universality extends to other situations, in particular to cold and dense systems. Some reason for optimism may be derived from the fact that the application of the duality to strongly correlated condensed matter systems has in recent years become a very active and successful field of research (see, e.g., \cite{Hartnoll:2009sz,McGreevy:2009xe,Adams:2012th} for reviews).

In the context of cold and dense QCD, the gauge/gravity duality has been applied to mimic the confined quark matter phase \cite{Bergman:2007wp,Rozali:2007rx,Kim:2007vd,Kim:2011da,Kaplunovsky:2012gb,Ghoroku:2013gja,Li:2015uea,Elliot-Ripley:2016uwb} as well as the deconfined quark matter phase \cite{Burikham:2010sw,Kim:2014pva,Hoyos:2016zke,Hoyos:2016cob,Ecker:2017fyh}. In \cite{Hoyos:2016zke} the analysis was further extended to the  description of NS matter, where a holographic equation of state (EoS) for quark matter was combined with state-of-the-art nuclear theory results from Chiral Effective Theory (CET) to construct a set of NS matter EoSs. The holographic result was seen to contain exactly one free parameter, $m_0$, corresponding to the three equal (constituent) quark masses, whose value was somewhat arbitrarily fixed to make the quark matter pressure vanish at the same baryon chemical potential as that of nuclear matter. This resulted in a strong first order deconfinement transition and the conclusion that the stars become unstable as soon as holographic quark matter begins to form inside their cores, so that no ``holographic HSs'' exist. It should, however, be noted that this approach neglected a number of important physical effects, including the differing bare masses of the quark flavors as well as quark pairing \cite{Alford:1997zt,Alford:1998mk,Alford:2007xm}, which has recently been approached using holography \cite{Faedo:2017aoe}. In addition, the holographic calculation was performed in the so-called probe brane limit, where the backreaction of the geometry to the presence of the branes is not taken into account. A significant improvement in the bottom-up holography was reported in \cite{Jokela:2018ers}, which led the authors to propose to-date the most realistic EoS for deconfined quark matter in the Veneziano limit.

In the paper at hand, we revisit the construction of holographic compact stars by combining the ``medium stiffness'' CET EoS of \cite{Hebeler:2013nza} with the quark matter EoS considered in \cite{Hoyos:2016zke}, but this time relaxing one of the assumptions made in the latter reference, namely the fixing of the parameter $m_0$ by the requirement that the pressures of the two phases vanish at the same chemical potential. This is seen to lead to a rich phenomenology, with the variation of $m_0$ generating four distinct types of compact stars. These include i) ordinary NSs, analogous to those constructed in  \cite{Hoyos:2016zke}; ii) pure QSs, composed of absolutely stable quark matter; and iii) and iv) hybrid stars of the second and third kind (the first kind referring to ordinary HSs), containing a nuclear matter core and either a quark matter crust (HS2), or a quark mantle and a nuclear matter crust (HS3). The viability of the solutions ii)-iv) is clearly subject to highly nontrivial assumptions about stable quark matter, and the physical nature of these star types is thus far from obvious. In particular, as discussed below, it should be noted that as our model only describes three-flavor quark matter, the stability of this phase as well as the nuclear matter one against two-flavor quark matter is an assumption of our calculation rather than a prediction thereof. Nevertheless, we feel that it is an interesting topic of research to study, how well the properties of these stars fit the available observational data on compact stars.

With the above caveats in mind, we find that for a range of values of $m_0$ our novel hybrid stars exhibit properties in excellent agreement with all known observational and theoretical bounds, including in particular their mass-radius relations and tidal deformabilities. Perhaps most interestingly, studying the tidal deformability of a 1.4 solar mass ($M_\odot$) star as a function of $m_0$, we find the quantity to be minimized not by ordinary NSs, but by an HS2 solution with a quark crust. As we shall explain below, this implies that our hybrid stars are in excellent agreement with the recent gravitational wave observation GW170817 of LIGO and Virgo \cite{TheLIGOScientific:2017qsa}. 

Our paper is structured as follows. In section 2, we introduce the holographic setup that we employ in the description of the quark matter phase, while section 3 contains details of the matching procedure of the nuclear and quark matter EoSs as well as an explanation of the qualitative properties of the different stellar solutions we discover. Section 4 is then dedicated to a more thorough comparison of the properties of these solutions with astrophysical (mainly LIGO) data and an inspection of the so-called universal relations, while conclusions are drawn in section 5. The appendices of the paper finally contain many important computational details concerning, e.g., the stability analysis of our star configurations, the derivation of analytic results for quark star solutions, and the determination of various astrophysical quantities that are used in section 4.

\section{Holographic model and setup}\label{sec:model}

The model we choose to describe quark matter with is based on $\cN=4$ $SU(N_c)$ supersymmetric Yang-Mills theory (SYM) with $N_f=3$ fundamental $\cN=2$ matter hypermultiplets that we treat in the quenched approximation and identify as the flavor fields  \cite{Karch:2002sh}. By introducing additional supersymmetry-breaking couplings in the theory, and integrating out the supersymmetric partners of gluons and quarks, this model can be continuously connected to QCD. As supersymmetry is also broken by the chemical potential, states with a large density could exhibit similar properties in the sectors that the two theories share, even though the theories have very different vacua. As discussed above, this is indeed what has been seen to happen at finite temperature and zero density.

The theory has a $U(1)$ axial symmetry that is explicitly broken when a nonzero mass is given to the flavor fields, in which case the chiral condensate is also non-zero. The mass is defined in a gauge-invariant way as the coefficient of the (supersymmetrized) bilinear quark operator $\bar{q}q$ in the renormalized action. In a slight abuse of language, we will refer to it as the ``quark mass''. Other definitions involving observables that are not gauge invariant (deriving for instance from the quark two-point function) cannot be computed using the holographic approach. Using the holographic dual description, it can be shown that the renormalized mass coincides with the energy gap between the vacuum and a state with a quark. For this reason, when matching to QCD, it is natural to consider the mass in the holographic model as closely related to the constituent quark mass, rather than to the bare quark mass.

To mimic finite quark density, we turn on a chemical potential for a $U_B(1)$ component of the global $U(N_f)\sim U(1)_B\times SU(N_f)$ flavor symmetry of the theory. For simplicity, we set  the quark masses to be all equal. \tred{Beta equilibrium and electric charge neutrality conditions are then automatically satisfied when the chemical potentials are equal to each other, $\mu_q\equiv\mu_B/N_c$. Clearly, this is just a rough approximation to QCD, and in particular we do not expect the model to capture all the details of the phase diagram with nonzero baryon density. Indeed, we restrict the application of the model to the EoS of flavor-symmetric deconfined matter, and in particular assume that it remains stable relative to two-flavor quark matter when the parameters of the model are extrapolated to fit QCD values}. Many improvements can be made on this approach, not only by introducing flavor-dependent masses, but for instance considering confining models that resemble QCD much more closely, such as the Sakai-Sugimoto model \cite{Sakai:2004cn}. The virtue of our model is, however, that it is the simplest and best studied holographic model, and, as we will show, it fares no worse than any other existing model when confronted with observations.

\subsection{Holographic description}

In the large-$N_c$ limit and at strong 't Hooft coupling $\lambda_{YM}\gg 1$, the $\cN=4$ SYM theory has a holographic description in terms of classical type IIB SUGRA in an $AdS_5\times S^5$ geometry \cite{Maldacena:1997re}. In the 't Hooft limit $N_f\ll N_c$, the flavor sector can be introduced as $N_f$ D7 probe branes extended along the $AdS_5$ directions and wrapping an $S^3\subset S^5$ \cite{Karch:2002sh}. At non-zero temperature $T$, the dual geometry is modified in the $AdS_5$ factor to a black brane
\be\label{eq:ads5metric}
ds^2=\frac{R^2}{r^2}\frac{dr^2}{f(r)}+\frac{r^2}{R^2}\left(-f(r)dt^2+dx_i^2\right)+R^2 d\Omega_5^2,\ \ f(r)=1-\frac{(\pi R^2 T)^2}{r^4},
\ee
where $R$ is the $AdS$ radius, related to the 't Hooft coupling through the string length $\sqrt{\alpha'}$, $\lambda_{YM}=R^4/(\alpha')^2$. When $T=0$ one recovers the usual $AdS_5$ metric in the Poincar\'e patch. In this case, a convenient set of coordinates is to combine the holographic radial direction $r$ with the $S^5$ directions into $\mathbb{R}^6\simeq  \mathbb{R}^4\times \mathbb{R}^2$ and use spherical coordinates for the $\mathbb{R}^4$ component
\be
dr^2+r^2 d\Omega_5^2=d\rho^2+\rho^2d\Omega_3^2+dy^2+dz^2.
\ee
Then, the $AdS_5\times S^5$ metric becomes
\be
ds^2=G_{MN}dX^MdX^N=\frac{\rho^2+y^2+z^2}{R^2}\eta_{\mu\nu}dx^\mu dx^\nu+\frac{R^2}{\rho^2+y^2+z^2}\left( d\rho^2+\rho^2d\Omega_3^2+dy^2+dz^2\right).
\ee
We will use the notation where $X^M$,  $M=0,1,\dots,9$ are the coordinates in the full ten-dimensional space, $x^\mu$, $\mu=0,\dots,3$ the coordinates along the field theory directions, $X^4=\rho$, $X^{5,6,7}$ the directions along the $S^3$ and $X^8=y$, $X^9=z$.

The flavor sector is mapped to probe D7 branes in the black brane background. The profile of the flavor branes in the background geometry is determined by the embedding functions $X^M(\sigma)$, depending on the worldvolume coordinates $\sigma^I$, $I=0,1,\dots,7$. The induced metric on the D7 brane worldvolume is
\be
g_{IJ}=G_{MN}(X) \partial_I X^M \partial_J X^N.
\ee
In addition, there is a $U(N_f)\sim U(1)_B\times SU(N_f)$ gauge field $A_I$ on the worldvolume of the D7 branes. We will restrict to Abelian configurations where only the $U(1)_B$ component is different from zero, and denote the field strength by $F_{IJ}=\partial_I A_J-\partial_J A_I$. The dynamics of the embedding functions and the D7 gauge field are determined by the classical action of the brane
\be
S_{D7}=-T_{D7}\int d^8 \sigma \,\sqrt{-\det\left( g_{IJ}+2\pi\alpha' F_{IJ}\right)}.
\ee
Although there could be an additional topological Wess-Zumino term, it vanishes for the background and configurations we are considering. The embedding is such that the worldvolume of the D7 lies along $AdS_5\times S^3$ directions
\be
X^\mu=\sigma^\mu,\ \ X^4=\rho=\sigma^4,\ \ X^{5,6,7}=\sigma^{5,6,7} .
\ee
The profile is given by
\be
X^8=y(\rho),\ X^9=z=0.
\ee
In addition, the time component of the $U(1)_B$ gauge field is allowed to depend on the radial coordinate $A_t(\rho)$. At zero temperature, the D7 brane action is
\be\label{eq:classicalD7action}
S_{D7}=-\frac{N_c N_f }{\lambda_\text{YM}}\frac{V_4}{(2\pi\alpha')^4}\int d\rho\, L(y',A_t'),\ \ L(y',A_t')=\rho^3\sqrt{1+(y')^2-(2\pi\alpha')^2 (A_t')^2}.
\ee
At non-zero temperature the black body factor $f(r)$ in \eqref{eq:ads5metric} forces us to work in a different set of coordinates and both the action and the solutions become more complicated, they have to be solved numerically or found by doing a perturbative expansion for small $T$.

\subsection{Thermodynamics}

In the absence of flavor, the free energy density can be obtained by evaluating the properly regularized classical SUGRA action on the black brane geometry. The result takes the form appropriate for a conformal field theory in four dimensions \cite{Gubser:1998nz}
\be
F_{\cN=4}=-\frac{\pi^2}{8}N_c^2T^4.
\ee
The thermodynamics of the model including flavor have been extensively studied in a number of previous works \cite{Mateos:2006nu,Kobayashi:2006sb,Mateos:2007vn,Karch:2007br,Mateos:2007vc,Erdmenger:2008yj,Ammon:2008fc,Basu:2008bh,Faulkner:2008hm,Ammon:2009fe,Erdmenger:2011hp, Jokela:2015aha,Itsios:2016ffv}. At zero temperature, there are two possible phases. At low chemical potential, the ground state has zero baryon density, and the meson spectrum is gapped as in the vacuum. When the chemical potential reaches a critical value equal to the quark mass, the theory, however, undergoes a phase transition to a state with non-zero baryon density and a gapless spectrum. In the phase with non-zero baryon density, the cost of introducing a quark is parametrically smaller (in the 't Hooft coupling) than in the gapped phase. The chiral condensate jumps through the phase transition, but remains non-zero as long as there is a non-zero quark mass.

In the dense phase, the free energy density naturally splits into two contributions,
\be
F=F_{\cN=4}+F_{\text{flavor}} \ ,
\ee
where only the latter part depends on the quark density. Being primarily interested in quiescent compact stars, we set the temperature to zero, in which case the $\cN=4$ part above vanishes, while the flavor part takes the simple analytic form \cite{Karch:2007br,Karch:2009eb,Itsios:2016ffv},\footnote{In principle this would take us beyond the regime of validity of the SUGRA approximation, as the backreaction of the D7 brane is not negligible at the horizon when the temperature is taken to zero, so the expression for the free energy should be taken as an extrapolation from small but nonzero temperatures.}
\begin{equation}\label{eq:freen}
F_{\text{flavor}}=-f_0(\mu_q^2-m_0^2)^2 \  .
\end{equation}
Here, we have defined $f_0=\frac{N_c N_f}{4 \gamma^3 \lambda_\text{YM}}$ with $\gamma=\Gamma(7/6)\Gamma(1/3)/\sqrt{\pi}$, while $m_0$ denotes the quark mass. This result can be obtained by solving the classical equations of motion for $y$ and $A_t$ derived from \eqref{eq:classicalD7action} and evaluating the action on-shell. Since the Lagrangian only depends on derivatives of the fields, there are two conserved quantities
\be
c=\frac{\partial L}{\partial y'},\ \ d=-\frac{1}{2\pi\alpha'}\frac{\partial L}{\partial A_t'}.
\ee
One can then solve algebraically for $y'$ and $A_t'$, giving
\be\label{eq:firstorderyAt}
y'=\frac{c}{\sqrt{\rho^6+d^2-c^2}},\ \ 2\pi \alpha' A_t'=\frac{d}{c}y'.
\ee
For $d=0$ and $c=0$, the embedding is constant $y=2\pi \alpha' m_0$ and $A_t=\mu$, thus it remains at a finite distance from the Poincar\'e horizon. This corresponds to the states with zero baryon density and a gapped spectrum. A quark is dual to a string extended between the horizon and the lowest point of the brane at $\rho=0$, thus the constituent mass is proportional to the length times the tension of the string $m_q=T_s y=m_0$.

For $d^2-c^2>0$, the embedding can be thought of as the zero temperature limit of D7 branes that reach the black hole horizon. The condition that the embedding reaches the horizon fixes the integration constant $y(0)=0$, and regularity at the horizon imposes $A_t(0)=0$. With these conditions, the solution is proportional to an incomplete Beta function
\be
y=\frac{1}{6}\frac{c}{(d^2-c^2)^{1/3}} B\left(\frac{\rho^6}{\rho^6+d^2-c^2}\,;\, \frac{1}{6}\,;\,\frac{1}{3} \right),\ \ 2\pi \alpha'A_t=\frac{d}{c} y.
\ee
The values at the asymptotic boundary can be identified with the mass of the quarks and the chemical potential, following the usual $AdS/CFT$ dictionary $y(\infty)=2\pi \alpha' m_0$, $A_t(\infty)=\mu_q$. This leads to the relations
\be
c=(2\pi\alpha')^3\gamma^{-3} (\mu_q^2-m_0^2) m_0, \ \ d=(2\pi\alpha')^3\gamma^{-3} (\mu_q^2-m_0^2) \mu.
\ee
The contribution of flavor to the free energy density is the D7 action  \eqref{eq:classicalD7action}  evaluated on the solution \eqref{eq:firstorderyAt}, minus a term that we subtract to remove the infinite volume divergence\footnote{This term can be understood as a local counterterm on the boundary, analogous to the counterterms that one has to introduce to renormalized field theories.}
\be
F_{\text{flavor}} =-\frac{1}{V_4}(S_{D7}-S_0)=\lim_{\Lambda\to \infty }\frac{N_c N_f }{\lambda_\text{YM}}\frac{1}{(2\pi\alpha')^4}\left[\int_0^{\Lambda} d\rho\frac{\rho^6}{\sqrt{\rho^6+d^2-c^2}}-\frac{\Lambda^4}{4}\right].
\ee
The result is
\be
F_{\text{flavor}} =-\frac{N_c N_f }{\lambda_\text{YM}}\frac{1}{(2\pi\alpha')^4}\frac{\gamma}{4}(d^2-c^2)^{2/3}=-\frac{N_c N_f }{4\gamma^3\lambda_\text{YM}}(\mu_q^2-m_0^2)^2.
\ee
At large values of the chemical potential, the effects of the mass are negligible and $F_{\text{flavor}}\sim \mu_q^4$ has the form expected for a conformal theory. Although this form is the same as for an ideal gas, one should note the dependence of the coefficient on the 't Hooft coupling. This is an indication that the theory remains strongly coupled at all values of the chemical potential.

The pressure $p$ and the energy density $\varepsilon$ are further determined from Eq.~\eqref{eq:freen} as 
\be
p=-F_{\text{flavor}}=f_0(\mu_q^2-m_0^2)^2, \ \ \varepsilon=\mu_q\frac{\partial p}{\partial \mu_q}-p=f_0(\mu_q^2-m_0^2)(3\mu_q^2+m_0^2),
\ee
which together lead to the EoS
\be\label{eq:eos}
\varepsilon  =  3p+4m_0^2\sqrt{f_0}\sqrt{p} \ .
\ee

It is worth stressing that an EoS of the above form (with a free energy as in Eq.~\eqref{eq:freen}) can also be obtained as a special case of the phenomenological model EoS of \cite{Alford:2004pf}. From this perspective, it might seem natural to extend the quark matter EoS by one further parameter, namely a constant representing the pressure difference between the confined and deconfined vacua, in analogy with the bag constant in the MIT bag model. We will, however, not study this possibility mainly because of the nature of the deconfinement transition within the large-$N_c$ holographic model. When the system moves from the gapped to the gapless phase, the free energy changes by an $O(N_c N_f)$ contribution, while the $O(N_c^2)$ part remains unaffected. This implies that the gapless phase describes finite charge density states within the original vacuum, so it is actually a relative of quarkyonic matter as introduced by McLerran and Pisarski in the context of large-$N_c$ QCD \cite{McLerran:2007qj}. For this same reason, we still regard $m_0$ as the constituent quark mass even after the transition to quark (or quarkyonic) matter has taken place.

The above result for the free energy is strictly valid in the large-$N_c$ limit, for fixed $N_f$ and very large 't Hooft coupling. We will assume that there are no significant changes when one moves away from this limit, in such a way that the qualitative behavior is correctly captured by Eq.~\eqref{eq:freen}. That is, the above EoS provides a zeroth order approximation to the physical system we wish to describe.

In our earlier work \cite{Hoyos:2016zke}, we fixed the parameters of the model to match the perturbative high-density limit of QCD, setting $N_c=N_f=3$ and $\lambda_\text{YM}=\frac{3\pi^2}{\gamma^3}\simeq 10.74$, while corrections entering with inverse powers of $N_c$ and $\lambda_\text{YM}$ were altogether neglected. In the present paper, we follow the same conventions, but let the parameter $m_0$ vary around the scale 310 MeV, where the nuclear matter pressure, chosen to follow the ``medium stiffness'' EoS of \cite{Hebeler:2013nza}, vanishes. This EoS corresponds to charge neutral beta-equilibrated matter and follows the CET result of \cite{Tews:2012fj} up to $1.1n_s$, thereafter extrapolating it with an observationally constrained piecewise polytropic form.

\section{Compact star solutions}\label{sec:structure}

As already noted, we construct NS matter EoSs by combining the medium stiffness nuclear matter EoS of \cite{Hebeler:2013nza} with a quark matter EoS obtained from the holographic model introduced in the previous section. At each value of the quark chemical potential, the phase that is realized is taken to be the one with lower free energy, or larger pressure, so that potentially there can be even multiple phase transitions inside the star. These will generically be of first order, with a latent heat that can be determined from the difference of the energy densities of the two phases at the transition.

\begin{figure}[t!]
\begin{center}
\includegraphics[width=0.45\textwidth]{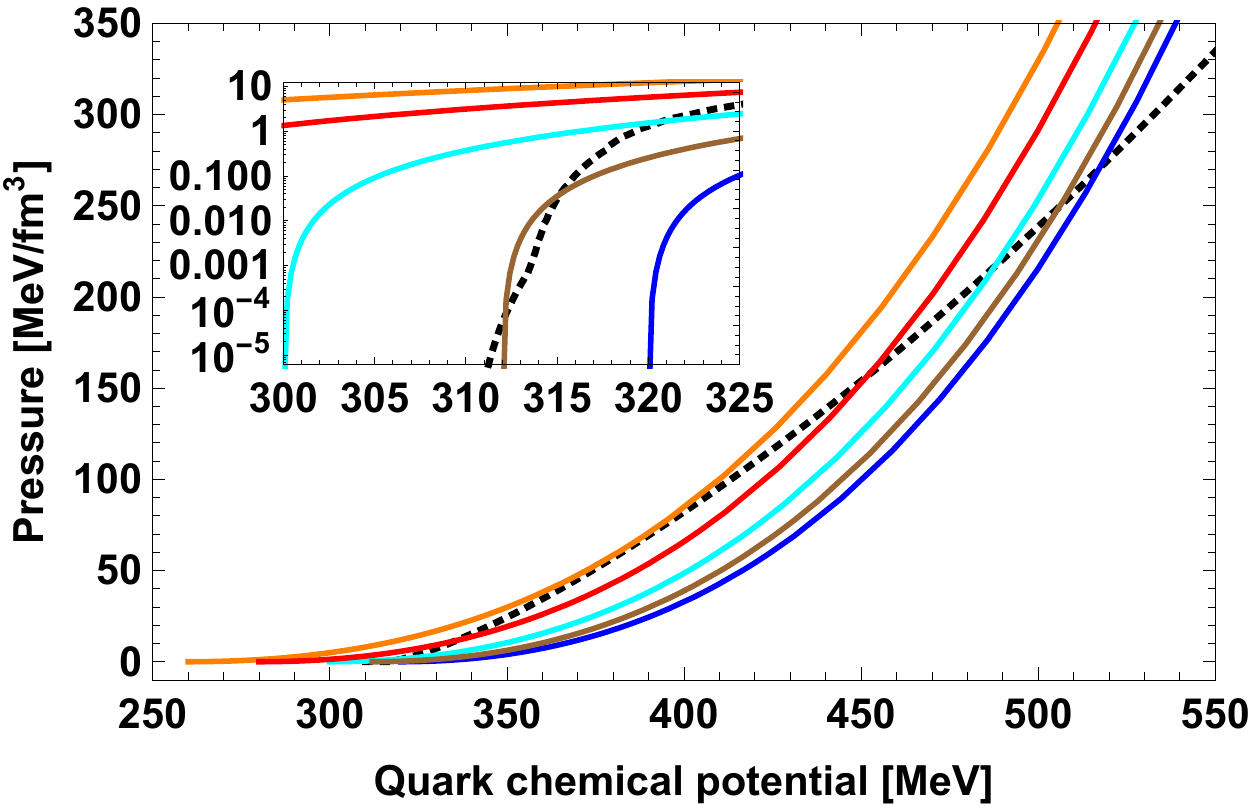}
\caption{The pressures of the nuclear and quark matter phases as functions of the quark chemical potential. The colored solid curves correspond to the holographic EoS for $m_0=260,280,300,312,320$ MeV (top-down), while the black dashed curve denotes the hadronic EoS taken from \cite{Hebeler:2013nza}.  The transitions happen at baryon densities between $n\sim 13-22\, n_s$ at high chemical potential and $n\sim 3\times 10^{-3}-0.15\, n_s$ at low chemical potential for the different curves shown, where $n_s\approx 0.16\, {fm}^{-3}$ is the saturation density.
}
\label{fig:Pvsmu}
\end{center}
\end{figure}

In Fig.~\ref{fig:Pvsmu}, we show the pressure of the nuclear matter phase together with that of the holographic one, giving $m_0$ the values 260, 280, 300, 312, and 320 MeV. These numbers have been chosen so that the cases displayed represent all of the four distinct scenarios we discover:
\begin{enumerate}
 \item For $m_0\gtrsim  313.1$ MeV, the nuclear matter pressure is dominant at low densities, but the quark matter phase takes over at a first order transition at some higher density, or chemical potential. 
 \item For $310.0\;{\rm MeV} \lesssim m_0\lesssim  313.1$ MeV, nuclear matter is still dominant at the lowest densities and quark matter at the highest, but between these regions there are not one but three first order transitions, so that counting from the lowest to the highest density, the phases of QCD matter are nuclear, quark, nuclear, and again quark matter.
 \item For $261.4\,{\rm MeV} \lesssim m_0 \lesssim310.0$ MeV, quark matter turns out to be favored both at the lowest and highest densities (i.e., it is stable in vacuum), but at moderate densities there exists a density interval where the nuclear matter pressure is larger.
 \item For $m_0\lesssim  261.4$ MeV, the pressure of quark matter is larger at all densities.
\end{enumerate}
The second and third of these scenarios are clearly nonstandard. Upon closer inspection, their existence can be traced back to the similar functional form of our holographic EoS for quark matter, Eq.~(\ref{eq:eos}), with that of the nuclear matter phase at low densities. For scenarios 3 and 4, we should in principle make sure that two-flavor quark matter is \textit{not} favored with respect to ordinary nuclear matter, but since our model is tuned to three quark flavors, we simply assume this to be the case.

\begin{figure}[t!]
\begin{center}
\includegraphics[width=0.45\textwidth]{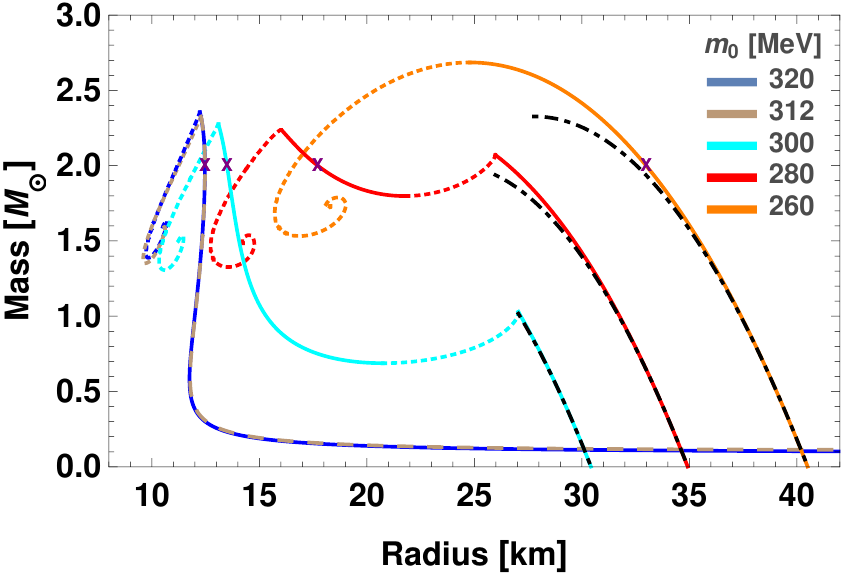}
\caption{$M$-$R$ curves for stellar solutions corresponding to the five values of $m_0$ displayed in Fig.~\ref{fig:Pvsmu}. The solid curves represent stable configurations, whereas the dashed ones illustrate unstable branches, assuming fast phase transitions, meaning that the phase changes instantaneously when the pressure fluctuates around the critical value \cite{Pereira:2017rmp}. Should this not be the case, the stability of some configurations may change. The dot-dashed black curves finally correspond to the analytic solutions of Eq.~(\ref{starmassnum}). Note that the 312 and 320 MeV curves lie practically on top of each other.
}\label{fig:MRmulti2}
\end{center}
\end{figure}

To study the properties of the compact stars built from the above EoSs ---  and to verify the claims made in the first section --- we next proceed to solve equations that govern relativistic hydrostatic equilibrium inside the stars. 
To this end, the metric of a spherically symmetric, non-rotating system can be written in the general form
\be
ds^2 =- e^{\nu(r)} c^2 dt^2 +\frac{dr^2}{  1 -\frac{2 G m}{c^2 r}}  + r^2 d\Omega_2 \ ,
\ee
while the structure of spherical stars in hydrostatic equilibrium can be solved from the TOV equations \cite{Oppenheimer:1939ne}
\be
\frac{dp}{dr} = \frac{G \left( \varepsilon + p \right)}{c^2 r^2}  \left( m + 4 \pi r^3 \frac{p}{c^2} \right) \left( 1 -\frac{2 G m}{c^2 r} \right)^{-1}  \ , \ \frac{dm}{dr} = 4 \pi r^2 \frac{\varepsilon}{c^2} \ , \ \frac{d\nu}{dr} = -\frac{2}{\varepsilon + p} \frac{dp}{dr}\ . \label{eq:tov}
\ee
Solutions to these equations can be found by specifying a value for the central pressure $p(r=0)$ and integrating the equations until the surface of the star, i.e.~the radius $R$ for which $p(r=R)=0$. Varying the central pressure finally produces a continuous curve on the mass-radius ($M$--$R$) plane, which specifies the possible masses and radii corresponding to the EoS studied. 

For each EoS, we not only solve the possible values of the stellar masses and radii, but in addition determine the stability of the configurations against infinitesimal adiabatic radial oscillations \cite{Chandrasekhar:1964zza,Chandrasekhar:1964zz}, assuming the transitions to be fast, the stability analysis is described in Appendix~\ref{app:stability}. The result of this exercise is shown in Fig.~\ref{fig:MRmulti2}, where $M$--$R$ curves corresponding to the five example EoSs of Fig.~\ref{fig:Pvsmu} are displayed.

Following the numbering of $m_0$ intervals introduced above, we now find:
\begin{enumerate}
 \item For $m_0\gtrsim  313.1$ MeV, the stars are always ordinary NSs, obeying an $M$--$R$ relation fully determined by the results of \cite{Hebeler:2013nza}. Quark cores are excluded by stability arguments due to a strong first order deconfinement transition (cf.~the discussion in \cite{Hoyos:2016zke}). 
 \item For $310.2\;{\rm MeV} \lesssim m_0\lesssim  313.1$ MeV, the stars are always of type HS3.
 \item For $264.4\,{\rm MeV} \lesssim m_0 \lesssim310.2$ MeV, two stable solutions exist: QSs at large and HS2s at small radii.
 \item For $m_0\lesssim  264.4$ MeV, all the stars are QSs.
 \end{enumerate}
Of particular interest here are clearly those HS2s and HS3s, for which $m_0$ is only slightly below the critical value of 313.1 MeV. Zooming into values of $m_0$ close to the critical one, we observe the $M$--$R$ relations to smoothly flow to that of ordinary NSs, just as expected. It is interesting to note that qualitatively similar solutions have been found earlier based on an MIT bag model EoS supplemented by a contribution from quark pairing \cite{Alford:2002rj,Alford:2004pf} (see also \cite{Zdunik:2012dj}).

Finally, let us note that for small compactness $C=GM/(c^2 R)$, we can analytically solve the TOV equations (\ref{eq:tov}) perturbatively if the EoS (\ref{eq:eos}) is assumed, i.e. for pure quark matter stars.
More specifically, the TOV equations can be solved in an expansion in a small parameter $\epsilon = (\mu_c-m_0)/m_0 \ll 1$, where $\mu_c$ is the central quark chemical potential of the star in question. For all other chemical
potentials, we then have $(\mu-m_0)/m_0 < \epsilon$, and it will also turn out that parametrically $C \sim \epsilon$.
To leading order in $\epsilon$ then, the EoS of the holographic model reduces to $\varepsilon \sim \sqrt{p}$, which corresponds to the Newtonian approximation of a fluid with a polytropic equation of state with adiabatic index $\gamma=2$. For such EoSs, an analytic solution to the TOV equations can be found in textbooks \cite{Glendenning:1997wn} even for general $\gamma$, resulting in
\be
R\sim M^{(\gamma-2)/(3\gamma-4)} \ .
\ee
For the special case of our $\gamma=2$, the radius is thus seen to be completely independent of the mass of the star. This relation is modified when corrections to the leading order result are taken into account.
To streamline the discussion, we have relegated this calculation in Appendix~\ref{app:ancal} and just state the final result for the $M$--$R$ relation:
\be\label{starmassnum}
M\simeq \frac{M_0}{c_0}\left[\frac{R_0-R}{R_0}-\frac{c_1}{c_0}\left(\frac{R-R_0}{R_0}\right)^2+\cdots\right] \ ,
\ee
where $c_0\simeq 1.853$, $c_1\simeq 2.948$, $R_0=\pi/k$, $M_0=c^2R_0/G$, and $k^2= 32\pi f_0 m_0^4 G/c^4$. We have included these analytical results for the QS in Fig.~\ref{fig:MRmulti2} as black dot-dashed curves and note that they match the numerics very accurately for small compactness.

\section{LIGO constraints and universal relations}\label{sec:universal}

It has been suggested long ago that in a coalescing binary system of two NSs, or a black hole and a NS, the tidal forces between the two objects affect the gravitational wave signal in a way that can be measured using Earth-based gravitational wave detectors \cite{Kochanek:1992,Bildsten1992,Kokkotas:1995xe,Taniguchi:1998jm,Pons:2001xs,Berti:2002ry,Mora:2003wt,Hansen:2005qv,Flanagan:2007ix}. In the fall of 2017, LIGO and Virgo were indeed able to place a quantitative limit on these effects in their analysis of gravitational wave data that very likely had their origins in the merger of two NSs \cite{TheLIGOScientific:2017qsa} (see also the analyses of \cite{Margalit:2017dij,Bauswein:2017vtn,Gupta:2017vsl,Zhou:2017xhf,Rezzolla:2017aly,Posfay:2017cor,Lai:2017mjv,Annala:2017llu,Radice:2017lry,Ayriyan:2017nby,Zhou:2017pha,Yang:2017gfb}). The limit was provided for the tidal deformabilities of the two stars involved in the merger --- a quantity related to the Love numbers of the stars that measures their susceptibility to the tidal forces that deform their shape. Importantly, these quantities are highly sensitive to the EoS of stellar matter, and it is thus of great interest to compute them for different candidate EoSs, including the ones introduced in our work. 

\begin{figure}[t]
\begin{center}
\includegraphics[width=0.45\textwidth,height=0.45\textwidth]{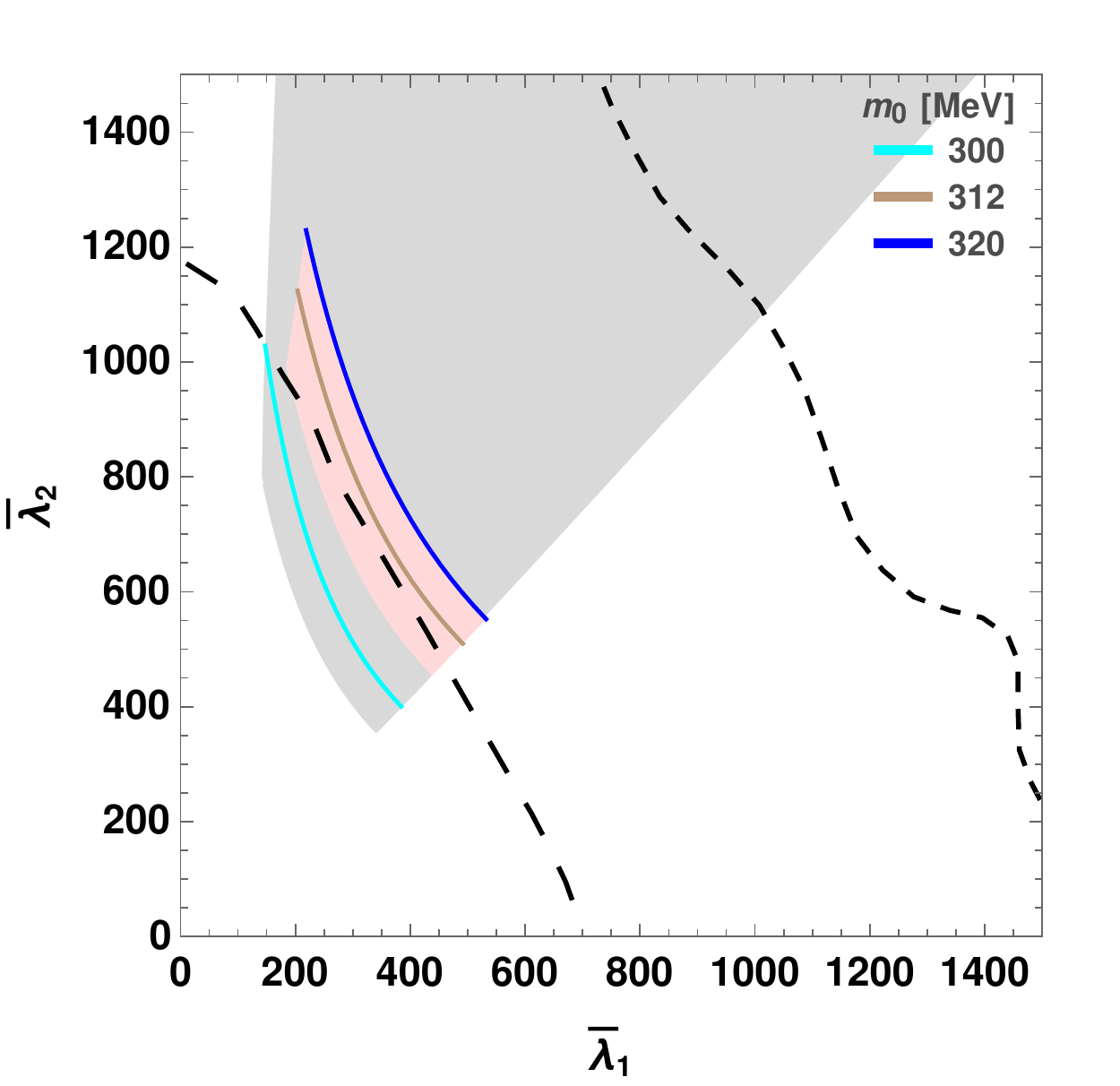}
\caption{The tidal deformabilities $\bar{\lambda}_i$ obtained for the two stars involved in the binary NS merger observed by LIGO and Virgo \cite{TheLIGOScientific:2017qsa}, corresponding to masses $M_1\in[1.36,1.60] \,M_\odot$ and $M_2\in[1.17,1.36] \,M_\odot$ (low-spin prior). The curves with the three different colors stand for the corresponding small-radius compact star solutions displayed in Fig.~\ref{fig:MRmulti2}. The curves corresponding to the remaining two EoSs of Fig.~\ref{fig:Pvsmu} fall outside the range of the plot. The gray area represents the set of all viable deformabilities obtained by varying $m_0$ for the type HS2 stars, whereas the type HS3 stars fall in the pink area.} \label{fig:obs}
\end{center}
\end{figure}

Another reason to be interested in Love numbers is that they allow the verification of so-called universal relations, i.e., suggested correlations between different quantities characterizing compact stars that appear to be largely insensitive to the EoS of stellar matter. These relations, due to Yagi and Yunes \cite{Yagi:2016bkt}, concern dimensionless ratios of the moment of inertia $I$, the quadrupolar moment of the mass distribution $Q$, and the electric Love number $k_2^{el}$ of compact stars,
\be
\bar{I} = \frac{c^4}{G^2 M^3} I, \ \ \bar{Q} = -\frac{M}{I^2} \frac{Q}{\Omega^2/c^2}, \ \ \bar{\lambda} = \frac{2}{3 C^5} k_2^{el} \ ,
\ee
where $\Omega$ is the angular velocity and $C$ the compactness of the star. It is clearly worthwhile to check, whether these relations hold for our family of EoSs as well.

\begin{figure}[t]
\begin{center}
\includegraphics[width=0.45\textwidth,height=0.3\textwidth]{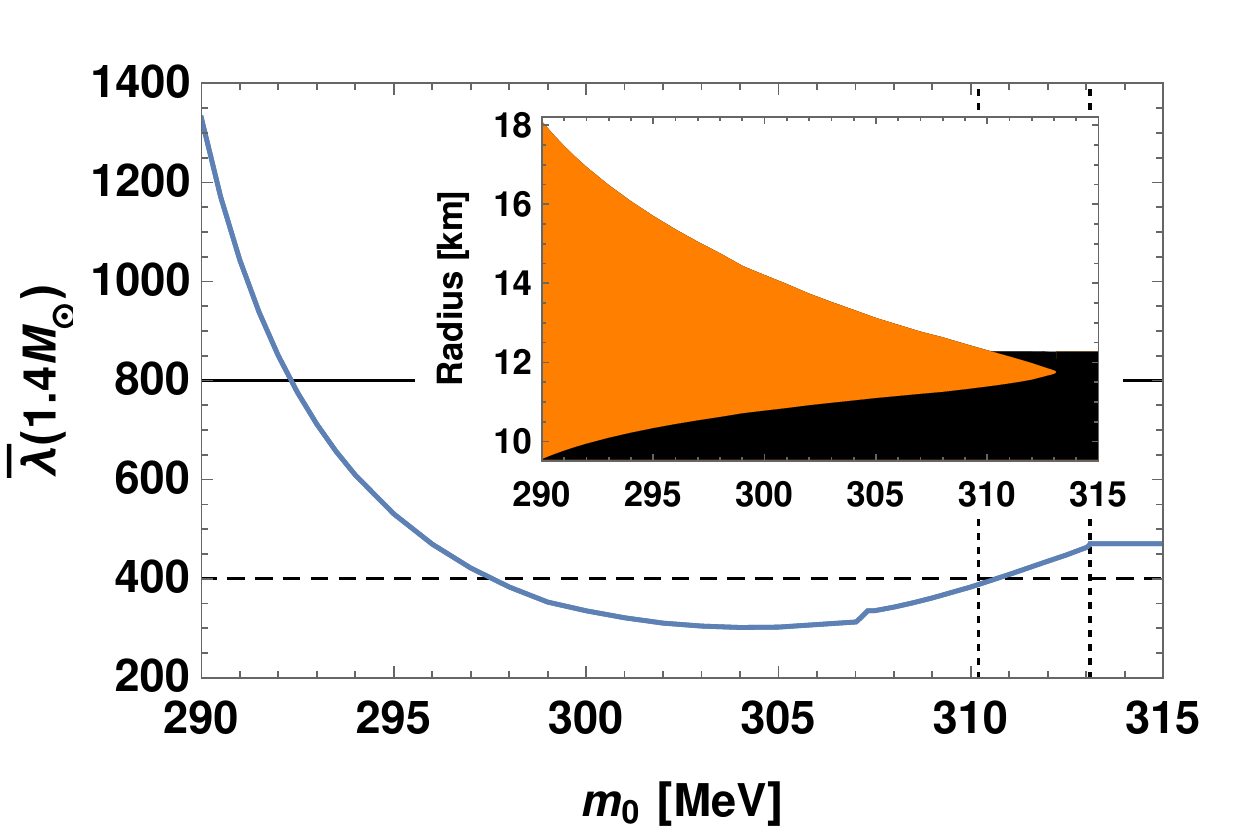}
\caption{The tidal deformability of a (lower-radius) $1.4M_\odot$ star as a function of $m_0$. Shown here are also horizontal lines denoting the values $\bar{\lambda}(1.4M_\odot)=800$ and 400, corresponding roughly to the 90\% and 50\% probability limits of LIGO and Virgo (cf.~discussion in \cite{Annala:2017llu}). The cusp in the curve around $m_0=307$ MeV is due to the matched EoS becoming sensitive a small discontinuity in the hadronic EoS of \cite{Hebeler:2013nza}. Inset: internal structure of hybrid stars of mass $1.4 M_\odot$, with the orange (black) color again representing quark (nuclear) matter. Vertical lines in the main plot indicate the transitions from HS2 to HS3 and from HS3 to NS as $m_0$ increases. The transition from QS to HS2 happens for lower values of $m_0$ than the ones shown in the plot.} \label{fig:obs2}
\end{center}
\end{figure}


Given a specific EoS, the determination of Love numbers involves perturbing the metric of a spherically symmetric (non-rotating) star with a quadrupolar deformation, as firstly introduced in \cite{Flanagan:2007ix,Hinderer:2007mb,Hinderer:2009ca}. These works were later generalized by including e.g. parity odd modes \cite{Binnington:2009bb,Damour:2009vw} and simplified \cite{Landry:2014jka} so that the master equation for $k_2^{el}$ can be written as:
\bea
r \eta' + \eta (\eta - 1) + A \eta - B = 0 \ ,
\eea
where
\bea
A &=& \frac{2}{f} \left[ 1 - 3 \frac{Gm}{c^2r} - 2 \pi \frac{G}{c^4} r^2 \left( \varepsilon + 3p \right) \right] \ , \\
B &=& \frac{1}{f} \left[ 6 - 4 \pi \frac{G}{c^4} r^2 (\varepsilon + p) \left( \frac{c^2}{c_s^2} + 3 \right) \right] \ ,
\eea
$c_s^2=\frac{\partial p}{\partial \varepsilon}$ is the speed of sound and $f = 1 - \frac{2Gm}{c^2r}$. At the center of the star $\eta(0) = 2$, and if we define $\eta_s \equiv \eta(R)$, then the matching condition at $r=R$ gives us \cite{Landry:2014jka}
\be
k^{\textup{el}}_2 = -\frac{1}{2} \frac{\eta_s - 2 - 4 C /  (1 - 2 C)}{\left[ \eta_s + 3 - 4 C / (1 - 2C) \right] B_1 - R B'_1} \ ,
\ee
where $B_1$ is a hypergeometric function:
\bea
B_1(r) &=&  {}_2F_1\left(3, 5; 6; \frac{2 G M}{c^2 r} \right) \ .
\eea
A more precise description can be found in Appendix~\ref{app:Love}.

At the same time, to obtain the quantities $I$ and $Q$ requires considering stars rotating with a small angular velocity $\Omega$. The moment of inertia $I$ can be obtained from the ratio of the angular momentum and the angular velocity \cite{Yagi:2013awa} which can be written as an ODE pair: \cite{Raithel:2016vtt,Glendenning:1997wn}
\be
\begin{split}
&\frac{dI}{dr} = \frac{8 \pi}{3 c^2} \frac{g j}{f} r^4 (\varepsilon + p) \ , \\
&\frac{d}{dr} \left(r^4 j \frac{dg}{dr} \right) + 4 r^3 \frac{dj}{dr} g = 0 \ ,
\end{split}
\ee
where 
\bea
j &\equiv& e^{-\nu/2} \sqrt{f} \ , \\
g &=& \frac{\tilde{\omega}}{\Omega}=1-\frac{\omega}{\Omega} \ ,
\eea
and $\omega$ is the angular velocity of the local inertial frame. By using the boundary conditions $g'(r=0)=0$ and $g(R)=1-2I/R^3$, the moment of inertia $I$ can be numerically determined. 

In case of $Q$, we must first determine the mass distribution inside the rotating star and then compute its second moment \cite{Yagi:2013awa}. As stated in \cite{Yagi:2013awa}, the essential interior solutions of the Einstein field equations in this particular case are
\bea
\frac{dK_2}{dr} &=& -\frac{dh_2}{dr} + \left(1 - 3 \frac{G m}{c^2 r} - 4 \pi \frac{G}{c^4} r^2 p \right) \frac{h_2}{r f} + \left(1 - \frac{G m}{c^2 r} + 4 \pi \frac{G}{c^4} r^2 p \right) \frac{m_2}{(r f)^2} \ , \\
\nonumber
\frac{dh_2}{dr} &=&  -\left(1 - \frac{G m}{c^2 r} + 4 \pi \frac{G}{c^4} r^2 p \right) \frac{1}{f} \frac{dK_2}{dr} + \left( 3 - 4 \pi \frac{G}{c^4} r^2 (\varepsilon + p) \right) \frac{h_2}{r f} + 2 \frac{K_2}{r f} \\
&& + \left( 1 + 8 \pi \frac{G}{c^4} r^2 p \right) \frac{m_2}{(r f)^2} + \frac{r^3 e^{-\nu}}{12 c^2} \left( \frac{d\tilde{\omega}}{dr} \right)^2 - \frac{4 \pi G r^3}{3 c^6 f} (\varepsilon + p) \tilde{\omega}^2 e^{- \nu} \ , \\
m_2 &=& - r f h_2 + \frac{r^4 f e^{- \nu}}{6 c^2} \left[ r f \left( \frac{d\tilde{\omega}}{dr} \right)^2 + 16 \pi \frac{G}{c^4} r (\varepsilon + p) \tilde{\omega}^2 \right] \ ,
\eea
and the corresponding Taylor expansion around the origin of the star:
\bea
h_2(r) &=& B r^2 + \mathcal{O}(r^4) \ , \\
K_2(r) &=& -B r^2 + \mathcal{O}(r^4) \ , \\
m_2(r) &=& -B r^3 + \mathcal{O}(r^5) \ ,
\eea
where $B$ is a constant related to the quadrupole moment. Besides, the corresponding exterior solutions can be given in rather simple forms: \cite{Yagi:2013awa}
\bea
h_2^{\text{ext}} &=& - \frac{3 A}{C (1-2C)} \left[ 1 - 3C + \frac{4}{3} C^2 + \frac{2}{3} C^3 + \frac{f(R)^2}{2C} \ln f(R) \right] + \left( \frac{L}{M R c} \right)^2 C \left( 1 + C \right) \ , \\
K_2^{\text{ext}} &=& \frac{3 A}{C} \left[ 1 + C - \frac{2}{3} C^2 + \frac{1-2C^2}{2C} \ln f(R) \right] - \left( \frac{L}{M R c} \right)^2 C \left( 1 + 2 C \right) \ , \\
m_2^{\text{ext}} &=&  \frac{3 A R}{C} \left[ 1 - 3C + \frac{4}{3} C^2 + \frac{2}{3} C^3 + \frac{f(R)^2}{2C} \ln f(R) \right] - \left( \frac{L}{M R c} \right)^2 C \left( 1 - 7 C + 10 C^2 \right) \ , 
\eea
where $L$ and $A$ are angular momentum and a matching constant. By insisting that the interior and exterior solutions of $K_2$ and $h_2$ match at the surface of the star, respectively, we can derive the values of constants $A$ and $B$. And by using above results, we can now calculate the quadrupole moment $Q$ of a star: \cite{Yagi:2013awa}
\be
Q = - \frac{L^2}{M c^2} - \frac{8}{5} \frac{G^2}{c^4} A M^3 \ .
\ee
More information about $I$ and $Q$ can be can be found in Appendix~\ref{app:IQ}.

We have performed numerically the calculations mentioned above for all the different types of compact stars we have encountered, and in addition provide analytic expressions at small compactness for the Love numbers in Appendix~\ref{app:pertlove} and for $I$ and $Q$, derived in detail in Appendix~\ref{app:IQ}. These leading order analytical results read 
\bea
\bar{I} & \simeq &\frac{2}{3} \frac{\pi^2-6}{\pi^2}\frac{1}{C^2} \simeq 0.261 C^{-2} \\
\bar{Q} & \simeq  &- \frac{32 \pi ^4 \left(15-\pi ^2\right) }{9+24 \pi  \left(\pi ^2-3\right)} C\simeq -30.35 C \  , 
\eea
together with $k_2^{el}\simeq 0.260-1.994 C$, which agree with our numerics to a good precision.


\begin{figure*}[t]
\begin{center}
\includegraphics[width=0.32\textwidth]{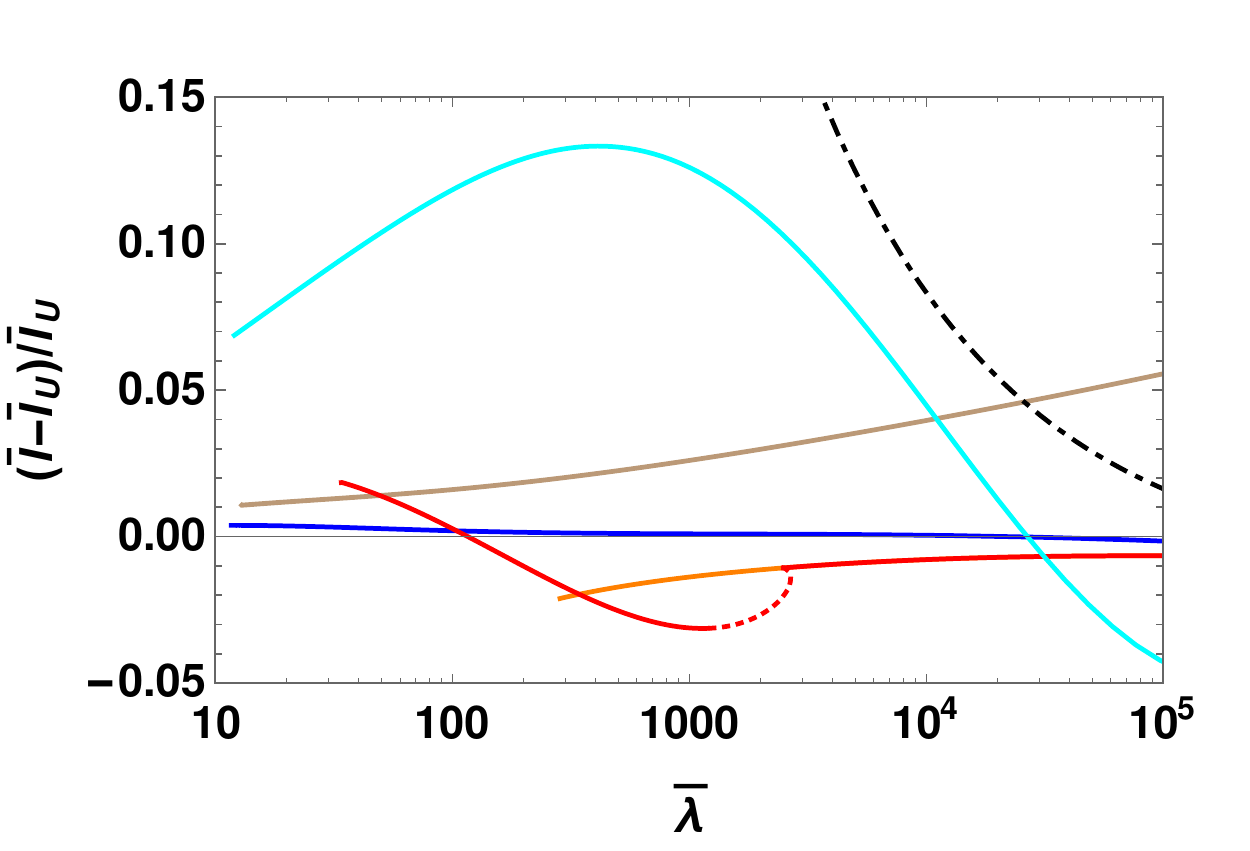}
\includegraphics[width=0.32\textwidth]{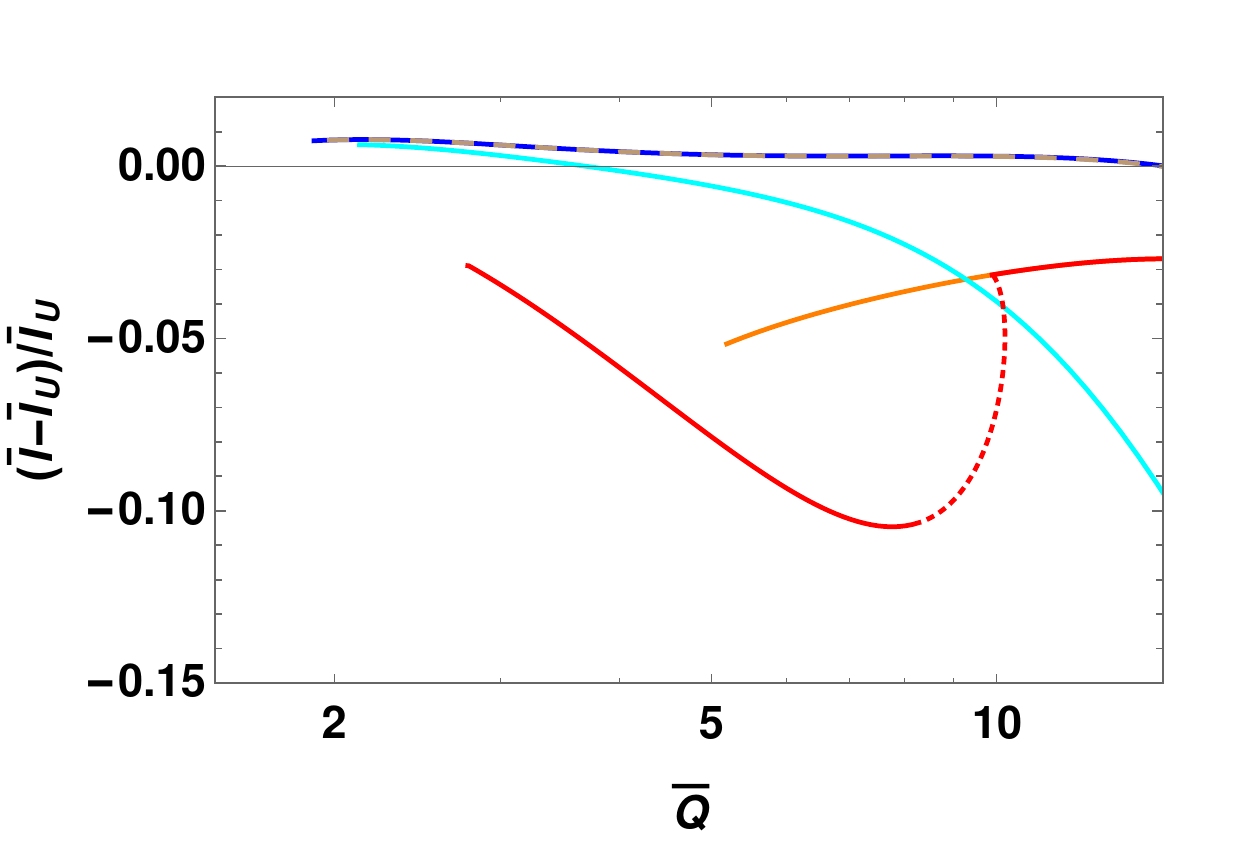}
\includegraphics[width=0.32\textwidth]{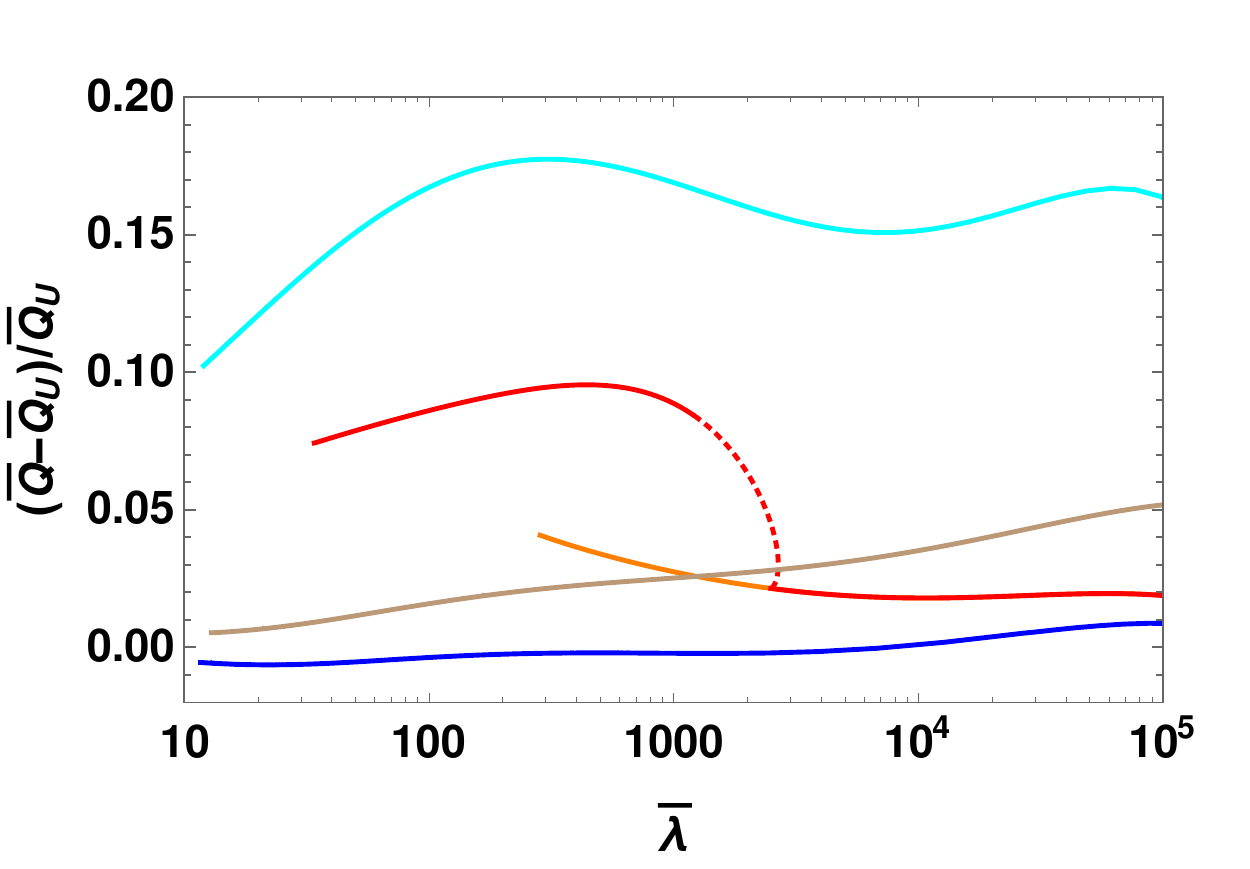}
\caption{The relative accuracy, to which the I-Love-Q universal relations of Yagi and Yunes \cite{Yagi:2016bkt} are reproduced by our five example EoSs. The subfigures present the $\bar{\lambda}$--$\bar{I}$ (left), $\bar{Q}$--$\bar{I}$ (middle), and $\bar{\lambda}$--$\bar{Q}$ (right) relationships, where the subscript $U$ always indicates that the value of the quantity is taken from the corresponding universal relation. In the $\bar{\lambda}$--$\bar{I}$ plot, the analytical curve is also shown as the black dot-dashed line, while for the other two cases this approximation falls outside the range chosen in the figure.} \label{fig:ILove}
\end{center}
\end{figure*}

Starting with the tidal deformability, we note that LIGO and Virgo provide the constraint $\bar{\lambda}(1.4M_\odot) \leq 800$ for the likely case of slowly rotating stars (the low-spin prior) at a 90\% Bayesian probability level \cite{TheLIGOScientific:2017qsa}. In addition to this, Fig.~5 of this reference gives both 90\% and 50\% probability contours for the independent tidal deformabilities of the two stars on a $\bar{\lambda}_1$-$\bar{\lambda}_2$ plane. To compare our results to these values, we first show in Fig.~\ref{fig:obs}, how our example EoSs from Fig.~\ref{fig:Pvsmu} relate to these contours. Here, the curves have been generated by independently determining the tidal deformabilities for both stars involved in the merger, obtaining the possible mass pairs by varying the mass of one of the two stars within the uncertainty region reported in \cite{TheLIGOScientific:2017qsa} and solving for the other using the accurately-known chirp mass of the event, $\mathcal{M}=1.188 M_{\odot}$. Interestingly, the smallest deformabilities are obtained not for ordinary NSs but for the HS2s and HS3s. 

To further inspect the rather surprising results observed, we next show in Fig.~\ref{fig:obs2} the tidal deformability value of a star of mass $1.4M_\odot$ as a function of $m_0$. Indeed, we verify from here that the quantity is minimized around $m_0=304$ MeV, i.e., for HS2s with a ca.~2 km thick quark crust (cf.~the inset of the figure). It is worth noting that the minimal value of $\bar{\lambda}(1.4M_\odot)=301$ is markedly smaller than that obtained for the NS solutions, $\bar{\lambda}(1.4M_\odot)=471$. This is very interesting to contrast with the recent claim of a lower bound existing for this quantity \cite{Radice:2017lry}.

Moving next on to the universal relations, we have quantitatively checked the relative accuracy, to which the I-Love-Q relations of Yagi and Yunes \cite{Yagi:2016bkt}, concerning the correlations $\bar{\lambda}$--$\bar{I}$, $\bar{Q}$--$\bar{I}$, and $\bar{\lambda}$--$\bar{Q}$, are reproduced by our compact stars corresponding to different $m_0$ values. Inspecting the results, depicted in Fig.~\ref{fig:ILove}, we find that the deviation from the universal limit is largest for the HS2 stars with relatively thick quark crusts, but quickly diminish as $m_0$ tends towards the critical value of 313.1 MeV. Although the deviations are never larger than 20\%, this finding may suggest a way of distinguishing the novel hybrid star solutions from the NS ones.

\section{Conclusions}\label{sec:conclusions}

As of today, holography remains the only computational tool that allows nonperturbative access to the properties of strongly coupled quantum field theories in those regions of parameter space where lattice methods are not applicable. While the holographic dual of QCD is still unknown, the strongly coupled regime of this theory covers practically all energies of phenomenological interest, including in particular the densities realized inside compact stars. Short of altogether circumventing the need to inspect the problematic density range by interpolating between trusted low- and high-density EoSs  \cite{Kurkela:2014vha,Annala:2017llu}, it is thus advisable to seek insights from novel directions, including theories whose strong-coupling limits can be reliably investigated using the gauge/gravity duality.

In the paper at hand, we have approached the description of moderate-density quark matter by studying a supersymmetric cousin of QCD. We derived a family of quark matter EoSs parameterized by the quark mass $m_0$, matched them with the EoS of beta-equilibrated nuclear matter \cite{Hebeler:2013nza}, and carefully applied the obtained results to the construction of compact stars. Taken at face value, our results suggest the possible existence of exotic hybrid stars, exhibiting features such as quark matter mantles or crusts. Interestingly, we found a range of values of $m_0$, for which these stars display both $M$--$R$ relations and tidal deformabilities in good agreement with available observational data. 

There are clearly a number of limitations in our approach, which range from the fact that the holographic model we study is not dual to QCD to the fact that we work in the so-called probe limit of the D3/D7 system, formally applicable only in the limit $N_f\ll N_c$. \tred{In addition, we needed to make nontrivial assumptions about the stability of two-flavor quark matter, and were left with a model, which is not applicable to addressing many detailed questions about the phase diagram, such as flavor symmetry-breaking or the chiral phase transition.} Recalling the difficulties that other field theory approaches face in the description of dense QCD matter, we believe it is nevertheless worthwhile to address the problem with holographic machinery. In this sense, our work should be viewed only as a first-order approximation, to be refined by future works improving and building on our model by e.g.~considering holographic models exhibiting stiffer EoSs \cite{Hoyos:2016cob,Anabalon:2017eri,Ecker:2017fyh}. In addition, it is worth noting that we have so far only reported on results concerning bulk thermodynamic observables, leaving the very interesting study of strongly coupled transport phenomena for the future.


\begin{acknowledgments}
We are grateful to David Blaschke for helpful comments as well as for pointing us to relevant existing literature, and to Kent Yagi for his advice concerning the analysis of the I-Love-Q relations. In addition, we thank Paul Chesler, Aleksi Kurkela, James Lattimer, and Joonas N\"attil\"a for useful discussions. E.A.~has been supported by the Finnish Cultural Foundation, and N.J.~and A.V.~by the Academy of Finland, grants no.~273545 and 1268023, as well as by the European Research Council, grant no.~725369. C.H.~and D.R.F.~have been supported by the Spanish grant MINECO-16-FPA2015-63667-P, and C.H.~in addition by the Ramon y Cajal fellowship RYC-2012-10370 and D.R.F.~by the GRUPIN 14-108 research grant from Principado de Asturias and by the C09 of SFB 1170 research grant by the DFG. C.E.~has finally been supported by the Austrian Science Fund (FWF), projects no.~P27182-N27 and DKW1252-N27. 
\end{acknowledgments}

\appendix

\section{Stability}\label{app:stability}

It has been stated that a non-rotating star is stable if it is stable against both radial oscillations and convection \cite{Detweiler1973}. The convection stability criterion for spherical star is \cite{Kovetz:1967, Schutz:1970}
\be
\left( \frac{dp}{d \varepsilon} \right)_{star} < \left( \frac{\partial p }{\partial \varepsilon} \right)_s\ ,
\ee 
where the left-hand side refers to the variation of energy density and pressure inside the star at different radial positions, while the right-hand side is the adiabatic speed of sound squared. Because we have assumed that our star is an isentropic system, then it is marginally stable against convection.

Chandrasekhar \cite{Chandrasekhar:1964zza, Chandrasekhar:1964zz} was the first who derived the condition for spherical star to be stable against infinitesimal adiabatic radial oscillations. If we introduce a radial displacement
\be
\Delta r = e^{\nu/2} u_n(r) e^{i \sigma_n t} / r^2 \ ,
\label{eq:delta_r}
\ee
his results can be express in a Sturm--Liouville form \cite{Glendenning:1997wn}
\be
\frac{d}{dr} \left(\Pi 	\frac{d u_n}{dr} \right) + \left(Q + \sigma_n^2 W \right) u_n = 0 \ ,
\label{eq:rad_osc_Chandr}
\ee
where
\be
\begin{split}
&\Pi = \frac{\gamma p}{r^2} e^{(\lambda + 3 \nu)/2} \\
&Q = -\frac{4}{r^3} \frac{dp}{dr} e^{(\lambda + 3 \nu)/2} - \frac{8 \pi}{r^2} e^{3(\lambda + \nu)/2} + \frac{1}{r^2 (\varepsilon + p)} \left( \frac{dp}{dr} \right)^2 e^{(\lambda + 3 \nu)/2} \\
&W = \frac{\varepsilon + p}{r^2} e^{(3 \lambda + \nu)/2}\ .
\end{split}
\ee
The variables $u_n$ and $\sigma_n^2$ are the amplitude, and eigenfrequency of the oscillation, respectively, and
\be
\gamma = \left( \frac{d \ln p}{d \ln n} \right)_s 
\ee 
is the varying adiabatic index. The boundary conditions of Eq. \eqref{eq:rad_osc_Chandr} are $u_n \sim r^3$ about the origin and $u_n'(R)=0.$ 

It can be shown that all eigenvalues $\sigma_n$ of the above Sturm--Liouville equation are real and they form a monotonically increasing sequence $\sigma_0^2 < \sigma_1^2 < \sigma_2^2 < \dots$, where $\sigma_0$ is the fundamental mode \cite{Shapiro:1983du}. Because the time dependence of the fluctuations is $e^{i \sigma_n t}$ (see Eq. \eqref{eq:delta_r}), a normal mode is unstable only if $\sigma_n^2 < 0$. Therefore, the whole configuration is stable if the fundamental mode is real, {\emph{i.e.}}, $\sigma_0^2 > 0$.

Numerically it is more efficient to use $\xi = \Delta r / r$ and $\eta = \Delta p / p$ instead of the radial displacement $\Delta r$ and the corresponding Lagrangian perturbation of the pressure $\Delta p$. Then the results of \cite{Chandrasekhar:1964zza, Chandrasekhar:1964zz} can be written as two first-order differential equations \cite{Chanmugam:1977}
\be
\begin{split}
&\frac{d\xi}{dr} =  -\frac{1}{r} \left( 3\xi + \frac{\eta}{\gamma} \right) - \frac{dp}{dr} \frac{\eta}{p + \varepsilon} \ , \\
&\frac{d\eta}{dr} = \xi \left[ \frac{\sigma^2}{c^2} e^{\lambda-\nu} \left( \frac{p + \varepsilon}{p} \right) r - \frac{4}{p} \frac{dp}{dr} - \frac{8 \pi G}{c^4} e^\lambda \left( p + \varepsilon \right) r + \left( \frac{dp}{dr} \right)^2 \frac{r}{p \left( p + \varepsilon \right)} \right] \\
&+ \eta \left[ -\frac{dp}{dr} \frac{\varepsilon}{p \left(p + \varepsilon \right)} - \frac{4 \pi G}{c^4} \left( p + \varepsilon \right) r e^\lambda \right]\ .
\end{split}
\label{eq:rad_osc_Chanm}
\ee
Demanding that $\xi'$ and $\eta'$ are regular, we get the boundary conditions
\bea
\label{eq:rad_osc_Chanm_center} \eta(0) &=& -3 \gamma(0)  \\
\eta(R) &=& \xi(R) \left[ \left( 1 -\frac{2 G M}{c^2 R} \right)^{-1} \label{eq:rad_osc_Chanm_surface} \left( -\frac{\sigma^2 R^3}{G M} - \frac{G M}{c^2 R} \right) - 4 \right] \ ,
\eea
if the eigenfunctions are normalized such that $\xi(0) = 1$. 

The oscillation equations \eqref{eq:rad_osc_Chanm} were numerically integrated starting from the center of the star with a trial value of $\sigma^2$ and given initial conditions. By using shooting method we determined the values of $\sigma^2$ which satisfied the boundary condition at the surface, Eq. \eqref{eq:rad_osc_Chanm_surface}. The fundamental mode frequency corresponds to the eigenfunction $\xi$ that has no nodes in range $r \in (0,R)$ \cite{Shapiro:1983du}.

\section{Analytic quark star solutions}\label{app:ancal}

In this appendix we will derive the analytical mass-radius relationship in Eq.~(\ref{starmassnum}). To avoid cluttering in the equations we will work in units with $G=c=1$. It is useful to first change the dependence of the solutions on the chemical potential to a dependence on the deviation of $\mu_q$ from the critical value $m_0$, by introducing a new variable
\be
\mn=\frac{\mu_q-m_0}{m_0} \ .
\ee
Then, the pressure and energy density of the holographic model can be written as
\bea
p(\mn) &=& \Lambda^4 \mn^2(\mn+2)^2 \ , \\
\varepsilon(\mn) &=& \Lambda^4 \mn(\mn+2)\left(3\mn^2+6\mn+4\right) \ ,
\eea
where 
\be
\Lambda^4=f_0 m_0^4.
\ee
The scaling symmetry of TOV equations $p\to a^2 p,\ \varepsilon\to a^2 \varepsilon,\ r\to r/a$ allows us to fix $4\pi G\Lambda^4/c^4=1$. The TOV equations then become 
\be
p'=-\frac{1}{r^2}\frac{(\varepsilon+p)(M+r^3 p)}{1-\frac{2M}{r}},\ \ M'=r^2\varepsilon, \ \ \nu'=-2\frac{p'}{p+\varepsilon} \ ,
\ee
where now
\bea
p(\mn) &=& \mn^2(\mn+2)^2 \\
\varepsilon(\mn) &=& \mn(\mn+2)\left(3\mn^2+6\mn+4\right) \ .
\eea

\subsection{Perturbative expansion}\label{sec:epsexpan}

Let us assume that the chemical potential at the center of the star $\mu_c$ is very close to the critical value $m_0$, in which case we can introduce an expansion parameter
\be
\epsilon=\mn_c=\frac{\mu_c-m_0}{m_0}\ll 1 \ ,
\ee
which, however, satisfies $\mn\leq \mn_c$. For a generic quantity $X$, we introduce the expansion
\be
X=\sum_n \epsilon^n \ord{X}{n} \ ,
\ee
obtaining for the $O(\epsilon^3)$ the pressure and $O(\epsilon^2)$  energy density:
\bea
\op &=& 0\ ,\\
\oe &=& 8\om\ ,\\
\opp &= & 4 (\om)^2\ , \\
\oee &= & 8\omm+  16  (\om)^2\ ,\\
\oppp &=&  8\om\omm+4(\om)^3 \ .
\eea

\paragraph{Leading order solution}

The TOV equations read in the Newtonian approximation
\be
\opp'=-\frac{1}{r^2}\oe\oM,\ \ \oM'=r^2\oe, \ \ \on'=-2\frac{\op'}{\oe} \ ,
\ee
or in terms of $\mn$,
\bea
\label{aneq1} r^2\om' &=& -\oM\ ,\\ 
\label{aneq2} \oM' & =& 8r^2\om\ ,\\
\label{aneq3} \on' &=& -2\om'\ .
\eea
Equation \eqref{aneq3} can be integrated to give
\be
\on=-2\om+\ord{\nu_\infty}{1}, \ \ \ord{\nu_\infty}{1}=-2\ord{M}{1}\oR^{-1}=-2\ord{C}{1} \ ,
\ee
where $M$ and $R$ are the mass and radius of the star and $C=M R^{-1}$ stands for its compactness. Taking a derivative of Eq.~\eqref{aneq1} and using \eqref{aneq2} yields now
\be
\left(r^2\om'\right)'+k^2 r^2\om=0,\ \ k^2=8 \ ,
\ee
from which we obtain, imposing the condition $\om(r=0)=1$, the solution
\be
\om=\frac{\sin(k r)}{kr} \ .
\ee

The radius of the star is determined by the point where the pressure vanishes, or $\om=0$, which at this order leads to
\be
\oR=\frac{\pi}{k} \ .
\ee
The mass function reads on the other hand
\be
\oM=\frac{1}{k}\left( \sin( k r)-k r \cos(k r)\right) \ ,
\ee
so that to the present order, the mass of the star is simply
\be
M\simeq \epsilon \oM\Big |_{r=\oR}=\epsilon \frac{\pi}{k}=\epsilon \oR \ ,
\ee
and thus the compactness 
\be
C=MR^{-1}\simeq \epsilon \ \Rightarrow \ \ord{C}{1}=1, \ \ord{\nu_\infty}{1}=-2 \ .
\ee

\paragraph{Next-to-leading order solution}

At NLO, the TOV equations become
\bea
\label{tovnlo1} \oppp' &=&-\frac{1}{r^2}\left( \frac{2(\oM)^2\oe}{r} +\oee\oM+\oMM\oe+\opp (r^3\oe+\oM)\right)\\ 
\label{tovnlo2} \oMM' &= &r^2\oee \\ 
\label{tovnlo3} \onn' &=&-2\left( \frac{\oppp'}{\oe}-\frac{\opp'}{(\oe)^2}(\oee+\opp)\right)\ .
\eea
The solution of Eq.~\eqref{tovnlo3} will not be necessary for our present calculation, so we will not try to solve it. Multiplying Eq.~\eqref{tovnlo1} by $r^2/\oe$, taking a derivative, and using Eq.~\eqref{tovnlo2}, we get rid of the explicit $\oMM$ dependence. Since all the $O(\epsilon)$ functions are known explicitly, we are left with an equation that can be solved analytically. After some algebra, one finds
\be
\left(r^2\omm'\right)'+k^2 r^2\omm+\ord{J}{2}=0 \ , 
\ee
where the inhomogeneous term reads
\be
 \ord{J}{2}=\frac{3 k^2}{2}\left[3 r^2(\om)^2+ 2 r\om \oM- \frac{2}{r^2}(\oM)^2\right] \ .
\ee
Imposing regularity, the solution to the inhomogeneous equation is
\be
\omm= \frac{1}{k^2 r^2} g(k r) \ ,
\ee
where we have defined
\be
g(z)=-\frac{1}{4} \Big\{2 z\Big[  3(\text{Ci}(z)-\text{Ci}(3 z)) \sin z+3 (\text{Si}(3 z)-3 \text{Si}(z)) \cos z +\sin (2 z)\Big]-6 \sin ^2 z\Big\} \ ,
\ee
and the $\cos$ and $\sin$ integrals are further defined as
\be
\text{Ci}(z)=-\int_z^\infty dt\,\frac{\cos t}{t}\ , \ \ \text{Si}(z)=\int_0^z dt\,\frac{\sin t}{t} \ .
\ee

The condition that the pressure vanishes at the surface of the star imposes $\mn=0$ at $r=R\simeq \oR+\epsilon \oRR$. Expanding in $\epsilon$, one finds from here
\be
\mn \simeq \epsilon \om+\epsilon^2 \omm+\cdots\simeq -\frac{\oRR}{\oR}\epsilon^2-c_0\epsilon^2+O(\epsilon^3) \ ,
\ee
where
\be
c_0=\frac{3}{2\pi}(3\text{Si}(\pi)-\text{Si}(3\pi))\simeq 1.85306 \ .
\ee
Therefore, the condition $\mn=0$ becomes 
\be\label{eq:R}
\oRR\simeq -c_0 \oR\ \Rightarrow \ \epsilon\simeq \frac{\oR-R}{c_0\oR} \ .
\ee

\subsection{Mass versus Radius formula}

To leading order, the mass decreases linearly with the radius,
\be
M\simeq \oR\epsilon \simeq \frac{1}{ c_0}\left( \oR-R\right) \ ,
\ee
which can be improved by considering the next order correction to the mass as well,
\be
\ord{M}{2}=\int_0^{\oR} dr r^2 \oee =\frac{1}{k}\int_0^\pi dz \left[2\sin^2 z+g(z)\right] \ , 
\ee
with a change of variable $z= kr$ in the second integral. The correction to the mass is proportional to the constant
\be
c_1\equiv -\frac{1}{\pi} \int_0^\pi dz \left[2\sin^2 z+g(z)\right]\simeq 2.948 \ ,
\ee
while the mass to this order becomes
\be
M\simeq \oR\left(\epsilon-c_1\epsilon^2\right) \ .
\ee
Using Eq.~\eqref{eq:R}, we can write the final result in the form
\be\label{starmass}
M\simeq \frac{1}{c_0}\left[\oR-R-\frac{c_1}{c_0\oR}\left(R-\oR\right)^2\right] \ .
\ee

\section{Love numbers}\label{app:Love}

We follow the conventions of \cite{Binnington:2009bb}. The metric is written in terms of the advanced lightcone coordinate $v$
\be
ds_0^2=-e^{2\psi} fdv^2+2e^{\psi}dv dr+r^3d\Omega_2 \ ,
\ee
where $f=1-2m/r$. In the exterior the metric is Schwarzschild's for $\psi=0$. The metric is perturbed $g_{\mu\nu}=g_{\mu\nu}^0+p_{\mu\nu}$ keeping the light-cone gauge condition $p_{r\mu}=0$. The perturbation has a multipole expansion in spherical harmonics. Denoting with $A,B$ the indices along the sphere directions, parity-even perturbations are
\be
p_{vv}=h_{vv}^{lm}(r) Y^{lm}, \ p_{vA}=j_v^{lm}(r) Y_A^{lm},\ \ p_{AB}=r^2 K^{lm}(r) \Omega_{AB}Y^{lm}+r^2 G^{lm}(r) Y_{AB}^{lm} \ ,
\ee
where $Y^{lm}$ are the usual scalar spherical harmonics, $\Omega_{AB}$ is the metric on the unit radius sphere and $Y_A^{lm}=D_A Y^{lm}$, $Y_{AB}^{lm}=\left( D_A D_B+\frac{1}{2}l(l+1)\Omega_{AB}\right)Y^{lm}$, with $D_A$ covariant derivatives on the sphere compatible with $\Omega_{AB}$. The parity-odd perturbations take the form
\be
p_{vA}=h_v^{lm}(r) X_A^{lm},\ \ p_{AB}=h_2^{lm}X_{AB}^{lm} \ ,
\ee
where the parity-odd vector and tensor harmonics are defined as $X_A^{lm}=-\epsilon_A^{\ B} D_B Y^{lm}$, $X_{AB}=-\frac{1}{2}\left(\epsilon_A^{\ C} D_B+\epsilon_B^{\ C} D_A \right)Y^{lm}$. We will work with the gauge-independent combinations defined as
\be
\tilde{h}_{vv}=h_{vv}+e^{-\psi}\left( e^{2\psi} f\right)' j_v-\frac{1}{2}r^2 f\left(e^{2\psi} f \right)' G', \ \ \tilde{h}_v=h_v \ .
\ee
The perturbation has two main contributions to the physics we want to describe. The first is an ``external'' quadrupolar deformation of the metric, that represents the contribution from an incoming gravitational wave. We are neglecting the time dependence and asymptotically the metric is not flat, so this should be taken as an approximation to a region around the star much smaller than the wavelength of a gravitational wave in the limit of small frequencies. The second contribution is due to the response of the matter in the star to the incoming wave. The matter distribution is modified and this, in turn, affects to the gravitational field surrounding the star. This is captured in the following expansion of the gauge-invariant variables in the region outside the star
\be
\begin{split}
&\tilde{h}_{vv}^{lm}=-\frac{2}{l(l-1)} r^l\left( 1+2 k_l^{\textup{el}}\left(\frac{R}{r}\right)^{2l+1}+\cdots\right){\cal E}^{lm}\ , \\
&\tilde{h}_{v}^{lm}=-\frac{2}{3l(l-1)} r^{l+1}\left( 1-2\frac{l+1}{l} k_l^{\textup{mag}}\left(\frac{R}{r}\right)^{2l+1}+\cdots\right){\cal B}^{lm} \ , 
\end{split}
\ee
where ${\cal E}^{lm}$, ${\cal B}^{lm}$ are polarization tensors. The coefficients $ k_l^{\textup{el}}$, $ k_l^{\textup{mag}}$ are the electric and magnetic Love numbers that characterize the response to the incoming gravitational wave.

For parity-even, or electric, gravitational Love number $k^{\textup{el}}_l$ the master equation is \cite{Binnington:2009bb, Landry:2014jka}
\be\label{eq:masterkel}
r^2 h''_{tt} + A h'_{tt} - B h_{tt} = 0 \ ,
\ee
where
\bea
A &=& \frac{2}{f} \left[ 1 - \frac{3m}{r} - 2 \pi r^2 \left( \varepsilon + 3 p \right) \right] \ , \\
B &=& \frac{1}{f} \left[ l(l + 1) - 4 \pi r^2 (\varepsilon + p) (c_s^{-2} + 3) \right] \ ,
\eea
and $f = 1 - 2 m / r$. We may simplify this equation by setting $\eta \equiv r h'_{tt}/h_{tt}$ then \cite{Landry:2014jka}
\be
r \eta' + \eta (\eta - 1) + A \eta - B = 0 \ .
\ee
The solution of the original master equation goes as $h_{tt} \propto r^l$ about the origin, which fixes $\eta(0) = l$. If we define $\eta_s \equiv \eta(R)$, then the matching condition at $r=R$ gives us \cite{Landry:2014jka}
\be
k^{\textup{el}}_l = \frac{1}{2} \frac{R A'_1 - \left[ \eta_s - l - 4 M /  (R - 2 M) \right] A_1}{\left[ \eta_s + l + 1 - 4 M / (R - 2M) \right] B_1 - R B'_1} \ ,
\ee
where $A_1$ and $B_1$ are hypergeometric functions:
\bea
A_1 &=& {}_2F_1\left(-l, 2-l; -2l; \frac{2M}{R} \right) \ , \\
B_1 &=&  {}_2F_1\left(l+1, l+3; 2l+2; \frac{2M}{R} \right) \ .
\eea

For odd-parity, or magnetic, gravitational Love number $k^{\textup{mag}}_l$ the corresponding master equation is \cite{Binnington:2009bb, Landry:2014jka}
\be\label{eq:masterkmag}
r^2 h''_t - P r h'_t - Q h_t = 0 \ ,
\ee
where
\bea
P &=& \frac{4 \pi r^2}{f} (\varepsilon + p)\ ,  \\
Q &=& \frac{1}{f} \left[ l (l + 1) -\frac{4 m}{r} + 8 \pi r^2 (\varepsilon + p) \right]\ .
\eea
As in the case of the electric Love number, we may write the corresponding gauge invariant metric perturbation as $\kappa \equiv r h'_t / h_t$. Then, the master equation has the form \cite{Landry:2014jka}
\be
r \kappa' + \kappa (\kappa-1) - P \kappa - Q = 0 \ ,
\ee
and $\kappa = l+1$ at the origin. Using the matching condition at the surface of the star, the magnetic Love number can be written as \cite{Landry:2014jka}
\be
k^{\textup{mag}}_l = \frac{l}{2 (l+1)} \frac{R A'_3 - (\kappa_s - l -1) A_3}{R B'_3 - (\kappa_s + l) B_3} \ ,
\ee
where $\kappa_s \equiv \kappa(R)$ and
\bea
A_3 &=&  {}_2F_1\left(1-l, -l-2; -2l; \frac{2M}{R} \right) \ , \\
B_3 &=&  {}_2F_1\left(l-1, l+2; 2l+2; \frac{2M}{R} \right) \ .
\eea

In \cite{Damour:2009vw} the surficial Love number in a relativistic setup was introduced. Ref. \cite{Landry:2014jka} even showed that there is an simple relation between this surficial Love number and the electric one:
\be
h_l = \Gamma_1 + 2 \Gamma_2 k^{\textup{el}}_l \ ,
\ee
where
\bea
\Gamma_1 &=& \frac{l+1}{l-1} \left(1 - \frac{M}{R} \right)  {}_2F_1\left(-l, -l; -2l; \frac{2M}{R} \right) - \frac{2}{l-1}  {}_2F_1\left(-l, -l-1; -2l; \frac{2M}{R} \right)  \\
\Gamma_2 &=& \frac{l}{l+2} \left(1 - \frac{M}{R} \right)  {}_2F_1\left(l+1, l+1; 2l+2; \frac{2M}{R} \right) - \frac{2}{l+2}  {}_2F_1\left(l+1, l; 2l+2; \frac{2M}{R} \right) \ .
\eea

\section{Perturbative calculation of Love numbers}\label{app:pertlove}

\subsection{Parity even modes}

For electric Love numbers, the problem reduces to solving the master equation \eqref{eq:masterkel}.
We will find analytic solutions at small compactness using the expansion
\be
\tilde{h}_{vv}=\oH+\epsilon \oHH+\cdots \ .
\ee
At each order in the expansion we have to solve an equation of the form
\be
r^2 \ord{H}{n}''+r \oA \ord{H}{n}'-\oB \ord{H}{n}+\ord{J}{n}=0 \ ,
\ee
where $\ord{J}{n}$ is determined by lower order solutions and coefficients, and $\ord{J}{0}=0$.
The coefficients in the master equation have the following expansion in the region inside the star $r\leq R$:
\be
\begin{split}
&\oAi=2 \ ,\\
&\oBi=l(l+1)- k^2 r^2 \ ,\\
&\oAAi=2\left[ -\frac{\ord{m}{1}}{r}-2\pi r^2\ord{(\varepsilon+3p)}{1}\right]\ , \\
&\oBBi=\left[2(l(l+1)-k^2 r^2)\frac{\ord{m}{1}}{r}-4\pi r^2 (3\ord{(\varepsilon+p)}{1} +\ord{(c_s^{-2}(\varepsilon+p))}{1}\right] \ .
\end{split}
\ee
Outside the star we have
\be
\begin{split}
&\oAo=2\ ,\\
&\oBo=l(l+1)\ ,\\
&\oAAo=-\frac{2\oM}{r}\ ,\\
&\oBBo=2l(l+1)\frac{\oM}{r} \ .
\end{split}
\ee
The first non-vanishing inhomogeneous term is
\be
\ord{J_i}{1}=r \oAAi\partial_r \oHi-\oBBi \oHi \ .
\ee

To leading order in the expansion, the inner and outer solutions are
\be
\begin{split}
&\oHi=\ord{c_i}{0} j_l(k r), \ \ r<R\ ,\\
&\oHo=\oao r^l+\obo r^{-l-1}, \ \ r>R \ .
\end{split}
\ee
Matching the two solutions at $r=R$ fixes
\be
\begin{split}
&\oao =\frac{\ord{c_i}{0}}{2l+1}R^{-l}\left[k R j_l'(k R)+(l+1)j_l(k R)\right]=\frac{\ord{c_i}{0}}{2l+1}\sqrt{\frac{\pi}{2} k R }R^{-l} J_{l-\frac{1}{2}}(kR) \ ,\\
&\obo =-\frac{\ord{c_i}{0}}{2l+1}R^{l+1}\left[ kR j_l'(k R)-l j_l(k R) \right]=\frac{\ord{c_i}{0}}{2l+1}\sqrt{\frac{\pi}{2} k R }R^{l+1} J_{l+\frac{3}{2}}(kR) \ .
\end{split}
\ee
At the next order, the inner and outer solutions are
\be
\begin{split}
&\oHHi=-j_l(kr)\int_0^{k r} \frac{dx}{\left(x j_l(x)\right)^2} \int_0^xdx_1 \,  j_l(x_1) \ord{J_i}{1}(x_1), \ \ r<R\ , \\
&\oHHo=\oaao r^l+\obbo r^{-l-1}+\oM\left(-(l+2)\oao r^{l-1}+(l-1)\obo r^{-l-2}\right), \ \ r>R \ .
\end{split}
\ee
The matching at this order gives the conditions
\be
\begin{split}
& \oaao+R^{-(2l+1)} \obbo=\alpha(R)\ , \\
& l \oaao-(l+1) R^{-(2l+1)} \obbo=\beta(R) \ ,
\end{split}
\ee
where
\be
\begin{split}
&\alpha(R)=\oao \oM R^{-1}\left((l+2)-(l-1)R^{-(2l+1)} \frac{\obo}{\oao} \right)\\
&-R^{-l}j_l(kR)\int_0^{k R} \frac{dx}{\left(x j_l(x)\right)^2} \int_0^xdx_1 \,  j_l(x_1) \ord{J_i}{1}(x_1)\ ,\\
&\beta(R)=\oao (l-1)(l+2)\oM R^{-1}\left(1+R^{-(2l+1)} \frac{\obo}{\oao} \right)\\
&-R^{-l} kR  j_l'(kR)\int_0^{k R} \frac{dx}{\left(x j_l(x)\right)^2} \int_0^xdx_1 \,  j_l(x_1) \ord{J_i}{1}(x_1)- \frac{R^{-l}}{(kR)j_l(kR)}  \int_0^{k R}dx_1 \,  j_l(x_1) \ord{J_i}{1}(x_1) \ .
\end{split}
\ee

\subsection{Parity odd modes}

For magnetic Love numbers, the problem reduces to solving the master equation \eqref{eq:masterkmag}.
We will find analytic solutions at small compactness using the expansion introduced in subsection \ref{sec:epsexpan}.
\be
\tilde{h}_{v}=\oH+\epsilon \oHH+\cdots \ .
\ee
At each order in the expansion we have to solve an equation of the form
\be
r^2 \ord{H}{n}''-r \oP \ord{H}{n}'-\oQ \ord{H}{n}+\ord{J}{n}=0 \ ,
\ee
where $\ord{J}{n}$ is determined by lower order solutions and coefficients, and $\ord{J}{0}=0$.
The coefficients in the master equation have the following expansion in the region inside the star $r\leq R$.
\be
\begin{split}
&\oPi=0\ , \\
&\oQi=l(l+1)\ , \\
&\oPPi=4\pi r^2\ord{(\varepsilon+p)}{1}\ , \\
&\oQQi=2\left[(l(l+1)-2)\frac{\ord{m}{1}}{r}+4\pi r^2 \ord{(\varepsilon+p)}{1}\right] \ .
\end{split}
\ee
Outside the star we have $P_o=0$ and
\be
\begin{split}
&\oQo=l(l+1) \ ,\\
&\oQQo=2(l(l+1)-2)\frac{\oM}{r}\ .
\end{split}
\ee
The first non-vanishing inhomogeneous term is
\be
\ord{J_i}{1}=-r \oPPi\partial_r \oHi-\oQQi \oHi\ .
\ee

To leading order in the expansion, the inner and outer solutions are
\be
\oHi=\oHo=\oao r^{l+1} \ .
\ee
At the next order, the inner and outer solutions are
\be
\begin{split}
&\oHHi=-r^{l+1}\int_0^{k r}dx\, x^{-2l-2}\int_0^x dx_1\, x_1^{l-1} \ord{J_i}{1}(x_1), \ \ r<R \ ,\\
&\oHHo=\oaao r^{l+1}+\obbo  r^{-l}- \oM \frac{l(l+1)-2}{l} \oao r^{l}, \ \ r>R \ .
\end{split}
\ee
The matching at this order gives the conditions
\be
\begin{split}
& \oaao+R^{-(2l+1)} \obbo=\alpha(R)\ , \\
& (l+1) \oaao-l R^{-(2l+1)} \obbo=\beta(R) \ ,
\end{split}
\ee
where
\be
\begin{split}
&\alpha(R)=\frac{l(l+1)-2}{l} \oao \oM R^{-1}-\int_0^{k R}dx\, x^{-2l-2}\int_0^x dx_1\, x_1^{l-1} \ord{J_i}{1}(x_1)\ ,\\
&\beta(R)=(l(l+1)-2)\oao \oM R^{-1}-(l+1) \int_0^{k R}dx\, x^{-2l-2}\int_0^x dx_1\, x_1^{l-1} \ord{J_i}{1}(x_1)\\
&-(k R)^{-2l-1}\int_0^{kR} dx_1\, x_1^{l-1} \ord{J_i}{1}(x_1) \ .
\end{split}
\ee

\subsection{Estimates for Love numbers}

We will estimate the values of Love numbers obtained from the analytic calculation.

\subsubsection{Electric Love numbers}

The electric love numbers are determined to NLO by the solutions found before
\be
\begin{split}
&\kel\simeq \frac{1}{2R^{2l+1}}\frac{\obo+\epsilon \obbo+\cdots}{\oao+\epsilon \oaao+\cdots}\simeq  \frac{1}{2R^{2l+1}}\frac{\obo}{\oao}\left(1+\epsilon \left[\frac{\obbo}{\obo}-\frac{\oaao}{\oao}\right]\right)+\cdots\\
&=\okl+\epsilon \okkl+O(\epsilon^2) \ .
\end{split}
\ee
The leading order contribution is
\be
\okl=\frac{1}{2R^{2l+1}}\frac{\obo}{\oao}\Big|_{R=R_0}=\frac{J_{l+\frac{3}{2}}(\pi) }{2J_{l-\frac{1}{2}}(\pi)} \ .
\ee
The subleading correction is
\be
\epsilon \okkl=\partial_R\left(\frac{1}{2R^{2l+1}}\frac{\obo}{\oao}\right)\Big|_{R=R_0}(R-R_0)+\epsilon \Delta \okkl \ ,
\ee
where
\be
\Delta \okkl=-\okl\frac{\oaao}{\oao}\left(1-\frac{1}{\okl}\frac{\obbo}{2 R^{2l+1}\oaao}\right)\Big|_{R=R_0} \ .
\ee
Here we can use ($n_l=\pi j_l'(\pi)+(l+1)j_l(\pi)$)
\be
\begin{split}
&\frac{\obbo}{2 R^{2l+1}\oaao}\Big|_{R=R_0}=\frac{1}{2}\frac{l\alpha(R_0)-\beta(R_0)}{(l+1)\alpha(R_0)+\beta(R_0)}\ ,\\
&\frac{\oaao}{\oao}\Big|_{R=R_0}=R_0^l\frac{(l+1)\alpha(R_0)+\beta(R_0)}{\ord{c_i}{0}n_l} \ .
\end{split}
\ee
The subleading correction has a contribution of the form
\be
\partial_R\left(\frac{1}{2R^{2l+1}}\frac{\obo}{\oao}\right)\Big|_{R=R_0}=\frac{\pi}{R_0} \partial_x\left(\frac{J_{l+\frac{3}{2}}(x) }{2J_{l-\frac{1}{2}}(x)} \right)_{x=\pi} \ .
\ee
The remaining contribution is
\be
\begin{split}
&\Delta \okkl=- \okl\frac{(l+1)\hat{\alpha}_0+\hat{\beta}_0}{n_l}\left(1-\frac{1}{2\okl}\frac{l\hat{\alpha}_0-\hat{\beta}_0}{(l+1)\hat{\alpha}_0+\hat{\beta}_0}  \right)\\
&=- \left[ \frac{1}{n_l}\left(\okl(l+1)-\frac{l}{2}\right)\hat{\alpha}_0+\frac{1}{n_l}\left( \okl+\frac{1}{2}\right) \hat{\beta}_0\right] \ ,
\end{split}
\ee
where
\be
\begin{split}
&\hat{\alpha}_0=\frac{n_l}{2l+1}\left((l+2)-2(l-1)\okl \right)-j_l(\pi)\hat{K}_1\ ,\\
&\hat{\beta}_0=\frac{n_l}{2l+1} (l-1)(l+2)\left(1+2\okl\right)-\pi  j_l'(\pi)\hat{K}_1- \frac{1}{\pi j_l(\pi)} \hat{K}_2 \ ,
\end{split}
\ee
and $K_1=\ord{c_i}{0}\frac{\delta\mu_c}{\mu_0} \hat{K}_1$, $K_2=\ord{c_i}{0}\frac{\delta\mu_c}{\mu_0} \hat{K}_2$ are defined as
\be
K_1=\int_0^{\pi} \frac{dx}{\left(x j_l(x)\right)^2} \int_0^xdx_1 \,  j_l(x_1) \ord{J_i}{1}(x_1), \ �\ K_2= \int_0^{\pi}dx_1 \,  j_l(x_1) \ord{J_i}{1}(x_1)\ .
\ee

\subsubsection{Magnetic Love numbers}

The magnetic Love numbers become nonzero only at NLO
\be
\begin{split}
\kmag\simeq -\frac{l}{l+1}\frac{1}{2R^{2l+1}}\frac{\obo+\obbo+\cdots}{\oao+\oaao+\cdots}\simeq  -\frac{l}{l+1}\frac{1}{2R^{2l+1}}\frac{\obbo}{\oao}+\cdots=\epsilon \ord{\kmag}{1}+O(\epsilon^2) \ ,
\end{split}
\ee
where
\be
\epsilon \ord{\kmag}{1}= -\frac{l}{l+1}\frac{1}{2R^{2l+1}}\frac{\obbo}{\oao}\Big|_{R=R_0} \ .
\ee
Here we can use
\be
\begin{split}
&\frac{\obbo}{2 R^{2l+1}\oao}\Big|_{R=R_0}=\frac{1}{2(2l+1)}\frac{(l+1)\alpha(R_0)-\beta(R_0)}{\oao} \ .
\end{split}
\ee
We obtain
\be
\begin{split}
&\ord{\kmag}{1}=-\frac{l}{l+1}\frac{1}{2(2l+1)}\left((l+1)\hat{\alpha}_0-\hat{\beta}_0\right) \ ,
\end{split}
\ee
where 
\be
\begin{split}
&\hat{\alpha}_0=\frac{l(l+1)-2}{l} -\hat{K}_1\ , \\
&\hat{\beta}_0=(l(l+1)-2)  -(l+1) \hat{K}_1-\pi^{-2l-1}\hat{K}_2\ ,
\end{split}
\ee
and $K_1=\oao\frac{\delta\mu_c}{\mu_0} \hat{K}_1$, $K_2=\oao\frac{\delta\mu_c}{\mu_0} \hat{K}_2$ are defined as
\be
K_1=\int_0^{\pi}dx\, x^{-2l-2}\int_0^x dx_1\, x_1^{l-1} \ord{J_i}{1}(x_1), \ �\ K_2= \int_0^{\pi} dx_1\, x_1^{l-1} \ord{J_i}{1}(x_1) \ .
\ee

\subsubsection{Approximate values}

Evaluating the integrals that appear in the formulas for the Love numbers, we can give a numerical estimate of their value to leading order in compactness. This is summarized in the table below.

\begin{table}[ht!]
\begin{center}
\begin{tabular}{|c|c|c|}
\hline $l$ & $\kel$ & $\kmag$  \\
\hline 2 & 0.260-1.994 C & 0.041 C\\
 3 & 0.106-1.047 C & 0.018 C\\
 4 & 0.060-0.720 C & 0.0094 C\\
 5 & 0.039-0.551 C & 0.0055 C\\
 6 & 0.028-0.448 C & 0.0035 C\\
 7 & 0.021-0.378 C & 0.0024 C\\
 8 & 0.016-0.327 C & 0.0017 C\\
 9 & 0.013-0.288 C & 0.0012 C\\
 10 & 0.011-0.258 C & 0.00091 C\\ \hline
\end{tabular}
\caption{Electric and magnetic Love number as function of compactness $C$.}\label{tablelove}
\end{center}
\end{table}

\section{Moment of inertia and quadrupolar momentum}\label{app:IQ}

When the star is rotating, the geometry and matter distribution are modified from the spherical shape. If the rotation is slow, one can expand in the angular velocity $\Omega$. To second order, the perturbed metric in an appropriate choice of coordinates takes the form
 \be
 \begin{split}
& ds^2=-e^\nu\left( 1+2\Omega^2 (h_0+h_2 P_2)\right)dt^2+\frac{1+\frac{2\Omega^2(\delta m_0+\delta m_2 P_2)}{r-2Gm}}{1-\frac{2G m}{c^2 r}} dr^2\\
&+ r^2\left[ 1+2\Omega^2(v_2-h_2) P_2\right]\left( d\theta^2+\sin^2\theta(d\phi-\omega dt)^2\right) \ ,
 \end{split}
 \ee
where $P_2\equiv P_2(\cos\theta)$ is the Legendre polynomial of order 2 and $\omega$ is the angular velocity of the local inertial frame. It will be convenient to define the relative angular velocity of the fluid respect to the inertial frame:
\be
g=\frac{\tilde{\omega}}{\Omega}=1-\frac{\omega}{\Omega} \ .
\ee

The moment of inertia $I$ can be computed integrating the following equations \cite{Raithel:2016vtt,Glendenning:1997wn}
\be\label{eq:Ieq}
\begin{split}
&\frac{dI}{dr} = \frac{8 \pi}{3 c^2} g j r^4 (\varepsilon + p) \left( 1 -\frac{2 G m}{c^2 r} \right)^{-1}\ , \\
&\frac{d}{dr} \left(r^4 j \frac{dg}{dr} \right) + 4 r^3 \frac{dj}{dr} g = 0 \ ,
\end{split}
\ee
where 
\be
j \equiv e^{-\nu/2} \sqrt{1 -\frac{2 G m}{c^2 r}} \ , 
\ee
with the boundary conditions $g'(r=0)=0$ and $g(R)=1-2I/R^3$.

The quadrupolar momentum is defined as the $O(1/r^3)$ correction to the Newtonian potential at large radius $r\to \infty$
\be
\Omega^2 h_2=\frac{Q}{r^3}+\cdots \ .
\ee
It can be computed by integrating the following equations \cite{Hartle:1968}
\be
\begin{split}
&\frac{d v_2}{dr} = -\frac{d\nu}{dr} h_2+\frac{\Omega^2}{c^2}\left(\frac{1}{r}+\frac{1}{2}\frac{d\nu}{dr} \right)\left[ -\frac{1}{3}r^3 \frac{dj^2}{dr}g^2+\frac{1}{6}j^2 r^4\left( \frac{dg}{dr}\right)^2\right]\ ,\\
&\frac{d h_2}{dr}=\left[ -\frac{d\nu}{dr}+\frac{G}{c^2}\frac{r}{r-2Gm/c^2} \left(\frac{d\nu}{dr} \right)^{-1}\left(\frac{8\pi}{c^2} (\varepsilon+p)-\frac{4m}{r^3} \right)\right] h_2\\
&-\frac{4 v_2}{r(r-2Gm/c^2)}\left(\frac{d\nu}{dr} \right)^{-1}+\frac{1}{6}\frac{\Omega^2}{c^2}\left[\frac{1}{2}\frac{d\nu}{dr} r-\frac{1}{r-2Gm/c^2}\left(\frac{d\nu}{dr} \right)^{-1} \right]r^3 j^2 \left( \frac{dg}{dr}\right)^2\\
&-\frac{1}{3}\frac{\Omega^2}{c^2}\left[ \frac{1}{2}r\frac{d\nu}{dr}+\frac{1}{r-2Gm/c^2} \left(\frac{d\nu}{dr} \right)^{-1} \right]r^2  \frac{dj^2}{dr} g^2 \ ,
\end{split}
\ee
with the boundary conditions $h_2=v_2=0$ at $r=0$ and $r\to \infty$.

\subsection{Analytic solutions}

The expansion of \eqref{eq:Ieq}  to leading order shows that the moment of inertia is of $O(\epsilon)$, where 
\be\label{eq:I1val}
I^{(1)}=\frac{2}{3}\frac{c^2}{G k^3}\int_0^\pi dz z^4 \om=\frac{2}{3}\frac{c^2}{G k^3}\pi(\pi^2-6) \ ,
\ee
thus (for $m_0\approx 310$ MeV)
\be
I\simeq 8.144\times 10^{46}\, C \; {\rm g}\, {\rm cm}^2 \ .
\ee
For the other functions we define the dimensionless mass parameter, angular velocity, and radial coordinate
\be
\hat{M}=\frac{G M k}{c^2}, \ \ \hat{\Omega}=\frac{\Omega}{ck}, \ \ z= k r \ .
\ee
We find
\begin{itemize}
\item Leading order solutions for $g$:
\be
\begin{split}
r \leq  R, & \ \ 1-g=\hat{\omega}_0+4\hat{M}(1-\hat{\omega}_0)\left(\frac{2}{\pi^2}+\frac{2z\cos z+(z^2-2)\sin z}{z^3} \right) \ ,\\
r > R, & \ \ 1-g=\hat{\omega}_0+\hat{\omega}_1\frac{\pi}{3}\left(1-\frac{\pi^3}{z^3}\right) \ ,
\end{split}
\ee
where
\be
\hat{\omega}_0=1-\frac{2GI}{c^2R^3}, \ \ \hat{\omega}_1= -4\hat{M}(1-\hat{\omega}_0)\frac{\pi^2-6}{\pi^3} \ .
\ee
Then, using \eqref{eq:I1val},
\be\label{eq:omega0}
1-\hat{\omega}_0=\frac{2 G I}{c^2 R^3}\simeq \frac{4(\pi^2-6)}{3\pi^2}C \ .
\ee
\item Leading order solutions for $h_2$:
\be
\begin{split}
r \leq  R, & \ \ \Omega^2 h_2=\hat{h}_0 j_2(z)-\hat{\Omega}^2(1-\hat{\omega}_0)^2\frac{z^2}{3}\ , \\
r > R, & \ \ \Omega^2 h_2=\frac{\hat{q}}{320 z^3}+\frac{3 \hat{M} \hat{q}}{2560 z^4}-\hat{\Omega}^2\frac{\pi ^8 \kappa ^2 (25 \hat{M}+56 z)}{1008 z^5}+O\left(\hat{M}^2\right) \ .
\end{split}
\ee
\item Leading order solutions for $v_2$:
\be
\begin{split}
r \leq  R, & \ \ \Omega^2 v_2=\frac{\hat{h}_0 \hat{M}}{2 z} \left(j_2(z) \left(\left(z^2+1\right) \sin z-z \cos z\right)+z j_1(z) (z \cos z-\sin z)\right)\\
&+\frac{2}{3}\hat{M}\hat{\Omega}^2 (1-\hat{\omega}_0)^2 z (\sin z-z \cos z), \\
r > R, & \ \ \Omega^2 v_2=\frac{\hat{M} \hat{q}}{5120 z^4}+\frac{\hat{M}^2 \hat{q}}{10240 z^5}-\hat{\Omega}^2\frac{\pi ^8}{24 z^4} \hat{\omega}_1^2 +O\left(\hat{M}^3\right) \ .
\end{split}
\ee
\end{itemize}
Matching the solutions at $r=R$ fixes
\be
\hat{h}_0=\frac{\pi ^4 (1+32 \pi ) (1-\hat{\omega}_0)^2 }{9+24 \pi  \left(\pi ^2-3\right)}\hat{\Omega}^2+O(\hat{M}),\ \ \hat{q}=-\frac{2560 \pi ^6 \left(\pi ^2-15\right) (1-\hat{\omega}_0)^2 }{9+24 \pi  \left(\pi ^2-3\right)}\hat{\Omega}^2+O(\hat{M}) \ .
\ee
The asymptotic expansion of the Newtonian potential $h_2$ is
\be
\Omega^2 h_2\simeq \frac{\hat{q}}{320 z^3} \ .
\ee
Then, the quadrupolar momentum is
\be
Q\simeq \frac{c^2}{G k^3}\frac{\hat{q}}{320} \ . 
\ee
In units of the angular velocity, this becomes (for $m_0\approx 310$ MeV)
\be
\frac{Q}{\Omega^2/c^2}\simeq \frac{c^2}{G k^5}\frac{8 \pi ^6 \left(15-\pi ^2\right) (1-\hat{\omega}_0)^2 }{9+24 \pi  \left(\pi ^2-3\right)}\simeq 1.692\times 10^{59}\, C^2\;{\rm g}\,{\rm cm}^4 \ ,
\ee
or, by using \eqref{eq:omega0}
\be
\frac{Q}{\Omega^2/c^2}\simeq \frac{G}{c^2 } k\frac{32 \pi ^3 \left(15-\pi ^2\right) }{9+24 \pi  \left(\pi ^2-3\right)}I^2 \ . 
\ee

\bibliographystyle{JHEP}

\bibliography{hybridholoJHEP_merge}

\end{document}